\documentclass[preprint,showpacs,preprintnumbers,amsmath,amssymb]{revtex4}
\pdfoutput=1
\usepackage{graphicx}
\usepackage[dvipsnames]{xcolor}
\usepackage{subfigure}
\usepackage{dcolumn}
\usepackage{bm}
\usepackage{orcidlink}
\usepackage{multirow, array} 
\usepackage{float} 
\newcommand{\be}{\begin{equation}}
\newcommand{\en}{\end{equation}}
\newcommand{\bea}{\begin{eqnarray}}
\newcommand{\ena}{\end{eqnarray}}
\usepackage[utf8]{inputenc}
\UseRawInputEncoding
\begin{document}
\title{Classical and Semiclassical Stability of Emergent Universes in Jordan-Brans-Dicke Theory}
\author{Pedro Labra\~{n}a \orcidlink{0000-0002-6390-8580}}
\email{plabrana@ubiobio.cl}
\affiliation{Departamento de F\'{i}sica, Universidad del
B\'{i}o-B\'{i}o, Casilla 5-C, Concepci\'on, Chile.\\
Centro de Ciencias Exactas, Universidad del B\'{i}o-B\'{i}o, Casilla 447, Chill\'an, Chile.}
\author{Juan Ortiz \orcidlink{0000-0003-4392-680X}}
\email{juang.ortizc@campusucc.edu.co}
\affiliation{Departamento de F\'{i}sica, Universidad del B\'{i}o-B\'{i}o, Casilla 5-C, Concepci\'on, Chile.\\
Systems Engineering Program, Faculty of Engineering, Universidad Cooperativa de Colombia, Monter\'{i}a 230002, Colombia.}
\date{\today}
%
%
%
\begin{abstract}

The Emergent Universe scenario is based on the assumption that the universe originates from a past-eternal Einstein static (ES) state, subsequently evolving toward an inflationary phase and a hot Big Bang era. Such models are appealing as they provide nonsingular and geodesically complete cosmological histories. However, it has been argued by Mithani and Vilenkin \cite{Mithani:2014jva, Mithani:2011en, Mithani:2012ii, Mithani:2014toa, Mithani:2015ona} that, even when the ES state is classically stable, certain models can admit semiclassical tunneling channels leading to quantum decay toward configurations of vanishing scale factor.

In this work, we investigate the classical and semiclassical stability of the ES regime in the context of Jordan--Brans--Dicke (JBD) theory. We analyze the structure of the Wheeler-DeWitt potential in minisuperspace and study representative semiclassical tunneling channels compatible with the Hamiltonian constraint. We show that, for suitable choices of the JBD potential and model parameters, the ES configuration can be robust against both classical perturbations and the specific semiclassical tunneling channels analyzed in this work. Our results indicate that the quantum instability discussed by Mithani and Vilenkin may be avoided within certain regions of parameter space, while leaving open the possibility of more general tunneling processes beyond the scope of the present analysis.

\end{abstract}

\maketitle

\section{Introduction}
\label{Int}

The standard cosmological model \cite{weinberg,peebles,kolb} and the inflationary paradigm \cite{Guth1,Albrecht,Linde1,Linde2} have successfully provided a consistent description of the large-scale structure and evolution of our universe. 
However, despite this success, fundamental questions remain open. In particular, it is still unclear whether the universe had a definite origin characterized by an initial singularity, or whether it could be past-eternal, avoiding a beginning in a physical sense.
In this context, spacetime singularity theorems developed within 
inflationary cosmology indicate that the universe is geodesically 
incomplete in the past \cite{Borde:1993xh,Borde:1997pp,Borde:2001nh,
Guth:1999rh,Vilenkin:2002ev}. These theorems show that null and 
timelike geodesics are generically past-incomplete in inflationary 
models, independently of any energy condition, provided that the 
Hubble parameter remains strictly positive on average along 
past-directed geodesics.
This result suggests that an initial singularity is difficult to avoid, even in inflationary scenarios.

The search for cosmological scenarios without initial singularities has motivated the development of the so-called Emergent Universe (EU) models \cite{Ellis:2002we,Ellis:2003qz}. In these models, the universe emerges from a past-eternal Einstein static (ES) state, subsequently entering an inflationary phase and evolving into a standard hot Big Bang cosmology. The EU models evade the geometrical assumptions of the singularity theorems
\cite{Borde:1993xh,Borde:1997pp,Borde:2001nh,Guth:1999rh,Vilenkin:2002ev},
thereby providing explicit examples of nonsingular inflationary universes.

The original EU models were formulated in the context of General Relativity \cite{Ellis:2002we,Ellis:2003qz}. 
However, in this framework the ES solution is known to be unstable under homogeneous perturbations, as first discussed by Eddington \cite{Eddington} and further analyzed in Refs.~\cite{Gibbons:1987jt,Gibbons:1988bm,Harrison:1967zz,Barrow:2003ni}. 
This instability implies that small perturbations drive the system away from the static state, thereby spoiling the Emergent Universe scenario.

One possible way to overcome this difficulty is to consider extensions of General Relativity. 
In this direction, several models have been proposed in which the ES universe can be classically stable \cite{Mulryne:2005ef, Mukherjee:2005zt, Mukherjee:2006ds, Banerjee:2007qi, Lidsey:2006md, Nunes:2005ra, Paul:2021lvb, Bengochea:2021jvt, Thakur:2021ufp, Ilyas:2020zcb, Paul:2020bje, Thakur:2021ryp, Hamil:2021ivf, Debnath:2020bno, Debnath:2019mkz, Paul:2019oxo, Li:2019laq, Martineau:2018isp, Olmedo:2018ohq, Cai:2018ebs, Labrana:2018ogv, Debnath:2017xcu, Ghose:2017xop, Atazadeh:2017xwe, Khodadi:2016gyw, Alesci:2016xqa, Ghosh:2016fue, Hadi:2016uaw, Sharif:2016hcx, Bhattacharya:2016env, Labrana:2016jmm, Rios:2016trs, Khodadi:2015fav, Marachlian:2015cia, delCampo:2015yfa, Guendelman:2015uca, Heydarzade:2015dba, Dutta:2015fha, Singh:2015nqn, Paul:2015eja, Gangopadhyay:2014jfa, Beesham:2014tja, Labrana:2014yta, Chervon:2014tra, Chakraborty:2014ora, Labrana:2013oca, Cai:2013rna, Rudra:2012mu, Cai:2012yf, Liu:2012ww, Singh:2012zzf, Chattopadhyay:2011fp, delCampo:2011mq, Debnath:2011qi, Paul:2011nw, Chakraborty:2010zzb, Mukerji:2010zz, Mukerji:2010zza, Wu:2009ah, Beesham:2009zw, Kan:2009ws, Paul:2009csp, Debnath:2008nu, Banerjee:2007sg, Labrana:2011np, Guendelman:2014bva}. 
In particular, in the context of Jordan--Brans--Dicke (JBD) theory, it has been shown that static solutions can be stable not only against homogeneous perturbations but also under anisotropic and inhomogeneous modes \cite{delCampo:2007mp,delCampo:2009kp,Labrana:2018bkw,Huang:2014fia,Cossio:2018xor}.

Despite this progress, it has been argued by Mithani and Vilenkin \cite{Mithani:2014jva,Mithani:2011en,Mithani:2012ii,Mithani:2014toa,Mithani:2015ona} that even classically stable static or oscillating universes may admit semiclassical tunneling channels leading to quantum decay toward configurations of vanishing radius. 
This observation raises the question of whether the emergent universe scenario can remain viable once semiclassical effects are taken into account.

In this work, we revisit this question within the framework of JBD theory. 
In particular, we consider ES  configurations supported by a JBD theory with a self interacting potential and matter content corresponding to a scalar field and analyze both their classical and semiclassical stability.
At the semiclassical level, we construct the Wheeler-DeWitt (WDW) equation in minisuperspace and study the structure of the associated effective potential. 
Special attention is devoted to the zero loci of the WDW potential, which mark the configurations of vanishing kinetic energy and play a central role in determining the admissible semiclassical dynamics. 
By analyzing representative tunneling channels within a WKB approximation, we show that, for suitable choices of the JBD potential, transitions toward configurations with vanishing scale factor can be strongly suppressed.

Our results indicate that the semiclassical instability identified in previous analyses is not universal and may be avoided in a class of EU models formulated within JBD theory.

The Jordan-Brans-Dicke theory \cite{Jbd} provides a natural framework for such scenarios, as it incorporates a dynamical gravitational coupling determined by a scalar field. 
Originally motivated by Mach's principle, this theory also arises in modern contexts such as supergravity, Kaluza-Klein theories and effective string actions \cite{Freund:1982pg,Appelquist:1987nr,Fradkin:1984pq,Fradkin:1985ys,Callan:1985ia,Callan:1986jb,Green:1987sp}.

Finally, we note that nonsingular and geodesically complete Friedmann-Robertson-Walker cosmologies, and in particular the role of positive spatial curvature, have received renewed attention in a series of recent works~\cite{Lesnefsky:2022fen, Kinney:2023urn, Geshnizjani:2023hyd, Easson:2024fzn, Easson:2024vpp, Moghtaderi:2025cns, Ferreira:2025zsy, Burwig:2025lyq, Kinney:2026abc, Easson:2026abc, Burwig:2026abc}. Since the Emergent Universe scenario is formulated in the closed ($k=+1$) sector, where positive spatial curvature supports the past-eternal Einstein static state, our Jordan-Brans-Dicke realization provides a concrete scalar--tensor example within this broader setting. The present analysis further addresses the semiclassical stability of such complete solutions, an issue of relevance to this class of models.

The paper is organized as follows. In Sect.~\ref{sec:H_JBD} we present the Hamiltonian formulation of the JBD theory within the ADM formalism and its reduction to minisuperspace. In Sect.~\ref{ES_JBD} we study static solutions and analyze their classical stability. In Sect.~\ref{sec:WdW} we construct the WDW equation and investigate the semiclassical dynamics, including the structure of the effective potential and representative tunneling channels. 
In Sect.~\ref{sec:Examples} we discuss specific examples. 
In Sect.~\ref{sec:ZERO_LOC} we analyze the zero loci of the WDW potential and their role in determining admissible tunneling endpoints. Finally, in Sect.~\ref{sec:conclu} we summarize our conclusions.

\section{Hamiltonian formalism for a Jordan-Brans-Dicke Theory}\label{sec:H_JBD}

We consider the following JBD action for a self-interacting potential and matter, given
by \cite{Green:2012oqa}
\begin{equation} \label{accion_JBD}
S=\int d^{4}x\sqrt{-g}\left[-\frac{1}{2}\Phi \mathcal{R}+
\frac{1}{2}\frac{\omega}{\Phi}\partial_{\mu}\Phi\partial^{\mu}\Phi-V(\Phi)+ \mathcal{L}_m(\psi)\right],
\end{equation}
where $\mathcal{L}_m(\psi)$ denotes the Lagrangian density of the matter content
\begin{equation}\label{density_L}
\mathcal{L}_m(\psi)=\frac{1}{2}\partial_{\mu}\psi\partial^{\mu}\psi-W(\psi),
\end{equation}
${\cal R}$ is the Ricci scalar curvature, $\Phi$ is the JBD field, $\omega$ is the JBD parameter, $V(\Phi)$ is the potential associated to the JBD field, $\psi$ is the inflaton field and $W(\psi)$ its effective potential.

Throughout this work we use units in which $8\pi G = c = \hbar = 1$, and we adopt the metric signature $(+,-,-,-)$ together with the curvature conventions of Ref.~\cite{Landau:1982dva}, namely ${\cal R}^{\rho}{}_{\sigma\mu\nu} = \partial_\mu \Gamma^{\rho}_{\nu\sigma} - \partial_\nu \Gamma^{\rho}_{\mu\sigma} + \Gamma^{\rho}_{\mu\lambda}\Gamma^{\lambda}_{\nu\sigma} - \Gamma^{\rho}_{\nu\lambda}\Gamma^{\lambda}_{\mu\sigma}$ and ${\cal R}_{\mu\nu} = {\cal R}^{\rho}{}_{\mu\rho\nu}$. With these conventions the Einstein-Hilbert term appears in the action with the sign displayed in Eq.~(\ref{accion_JBD}), and a healthy scalar field has a kinetic term of the form $+\frac{1}{2}\partial_\mu\psi\,\partial^\mu\psi$, as in Eq.~(\ref{density_L}).

We use the minisuperspace approximation \cite{Vilenkin:1987kf}, which is appropriate for our model, where we are going to consider that the universe is homogeneous, isotropic and closed during the ES regime, see \cite{delCampo:2007mp, delCampo:2009kp}. Then, the metric is given by the following expression
\begin{equation}
ds^{2}=N^{2}(t)dt^{2}-a(t)^{2}\left[\frac{dr^{2}}{1-r^{2}}+r^2d\theta^{2}+r^{2}\sin^{2}
\theta d\phi^{2}\right ]\label{metrica_RW}.
\end{equation}
By evaluating the JBD action, Eq.~(\ref{accion_JBD}), in the metric Eq.~(\ref{metrica_RW}) we obtain the following Lagrangian
\begin{equation}
\begin{split}
L & =-2 \pi ^2 a^3 N W(\psi)-2 \pi ^2 a^3 N V(\Phi )+\frac{\pi ^2
a^3 \dot{\Phi }^2 \omega }{N \Phi }+\frac{\pi ^2 a^3 \dot{\psi
}^2}{N}-\frac{6 \pi ^2 \left(a^2 \dot{a}
\dot{\Phi }\right)}{N}\\
&\quad +6 \pi ^2 a N \Phi -\frac{6 \pi ^2 \left(a \dot{a}^2 \Phi \right)}{N}.\\
\end{split}
\end{equation}

The Hamiltonian of the model is given by
\begin{equation}
H=P_a \dot{a} + P_\Phi  \dot{\Phi }+P_\psi  \dot{\psi }-L \,\,,
\end{equation}
where the conjugate momenta are given by
\begin{equation}
P_{a}=\frac{\partial L}{\partial \dot{a}} =-\frac{6 \pi ^2 a\left(2
\dot{a} \Phi +a \dot{\Phi }\right)}{N},\label{119}
\end{equation}
\begin{equation}
P_{\Phi}=\frac{\partial L}{\partial \dot{\Phi}}=\frac{2 \pi ^2
a^2\left(a \dot{\Phi } \omega -3 \dot{a} \Phi \right)}{N \Phi
},\label{120}
\end{equation}
\begin{equation}
P_{\psi}=\frac{\partial L}{\partial \dot{\psi}}=\frac{2 \pi ^2 a^3 \dot{\psi }}{N},\label{PPsi}
\end{equation}

then, we obtain

\begin{eqnarray}\label{eq:hamiltonian}
H&=&-\frac{N\,\omega}{12\pi^2 (3+2\omega)\, \Phi\, a}\Big[P_a^2 - \frac{6\Phi^2 P_\Phi^2}{\omega a} + \frac{6\Phi P_a P_\Phi}{\omega a} -\frac{3\Phi P^2_\psi
(3+2\omega)}{\omega a^2} + \frac{72\pi^4 \Phi^2 (3+2\omega) a^2}{\omega}\nonumber\\
\label{hamiltoniano_1}\\\nonumber &&-\frac{24 \pi^4 \Phi a^4
(3+2\omega) V(\Phi)}{\omega} -\frac{24\pi^4 \Phi a^4 (3+2\omega)
W(\psi)}{\omega}\Big] = N \mathcal{H}\,.
\end{eqnarray}

The classical Hamiltonian constraint is $\mathcal{H}=0$. The
classical field equations are given by, $\dot{P}_a = -\frac{\partial \mathcal{H}}{\partial
a}$, $\dot{P}_\Phi = -\frac{\partial \mathcal{H}}{\partial \Phi}$ and
$\dot{P}_\psi = -\frac{\partial \mathcal{H}}{\partial \psi}$, which we can write as follows
\begin{equation}
\left(\frac{\dot{a}}{a}\right)^{2}+\frac{1}{a^{2}}+\frac{\dot{a}}{a}\,\frac{\dot{\Phi}}{\Phi}=
\frac{\rho}{3\Phi}+\frac{\omega}{6}\left(\frac{\dot{\Phi}}{\Phi}\right)^{2}\! +
\frac{V}{3\Phi}\label{E_campo_1},
\end{equation}
\begin{equation}
2\frac{\ddot{a}}{a}+H^{2}+\frac{1}{a^{2}}+
\frac{\ddot{\Phi}}{\Phi}+2H\frac{\dot{\Phi}}{\Phi}+\frac{\omega}{2}\left(\frac{\dot{\Phi}}{\Phi}\right)^{2} \!-
\frac{V}{\Phi}=-\frac{P}{\Phi}\label{17},
\end{equation}

\begin{equation}
\ddot{\Phi}+3H\dot{\Phi}=\frac{\rho-3P}{2\omega+3}+\frac{2}{2\omega+3}[2V-\Phi
V']\label{18},
\end{equation}

\begin{equation}
\ddot{\psi } + 3 H \dot{\psi }=-\frac{\partial W(\psi) }{\partial
\psi }\label{con_energia},
\end{equation}
where $\rho=\frac{\dot{\psi}^{2}}{2}+W(\psi)$, $P=\frac{\dot{\psi}^{2}}{2}-W(\psi)$ and $V'=\frac{d V(\Phi)}{d \Phi}$.

The equations (\ref{E_campo_1})-(\ref{con_energia}), determine the classical evolution of the JBD model described by the action Eq.~(\ref{accion_JBD}).

\section{Static Universe solution and its classical stability}\label{ES_JBD}

In this section we review, by following a Hamiltonian approach, the static universe solution discussed in
Refs.~\cite{delCampo:2007mp, delCampo:2009kp}. 

The static universe solution in a JBD theory is characterized by the conditions $a = a_0= Constant$,
$\dot{a}_0=0=\ddot{a}_0$ and $\Phi=\Phi_0=Constant$,
$\dot{\Phi}_0=0=\ddot{\Phi}_0$, see Ref.~\cite{delCampo:2007mp}.

Following the scheme of the model in Ref.~\cite{delCampo:2007mp}, we are going to consider that the matter potential $W(\psi)$ is flat, that is $W(\psi)=W_0=Constant$. This is justified, since usually in the EU models it is considered that during the static regime the inflaton field $\psi$ is rolling in the flat section of its potential, see 
Ref.~\cite{delCampo:2007mp}. 
Then, we can notice that during the static regime the momentum $P_\psi$ is conserved. Then we can write the Hamiltonian constraint as follows
\begin{eqnarray}\label{constrainH}
\mathcal{H} = -\frac{\omega}{12\pi^2 (3+2\omega)\, \Phi\,
a}\Big[P_a^2 -\frac{6\Phi^2 P_\Phi^2}{\omega a} + \frac{6\Phi P_a
P_\Phi}{\omega a}+ U(a,\Phi)\Big] = 0,
\end{eqnarray}
where we have defined the following effective potential

\begin{equation}
\begin{split}
U(a,\Phi) & =-\frac{3\, P_\psi^2\, \Phi  (2 \omega +3)}{a^2 \omega }+
\frac{72 \pi ^4 a^2 \Phi ^2 (2 \omega +3)}{\omega }-\frac{24 \pi ^4 a^4 W_0 \Phi
(2 \omega +3)}{\omega }\\ 
&\quad-\frac{24 \pi ^4 a^4 \Phi  (2 \omega +3) V(\Phi )}{\omega}.\label{Potencial(a,Ph)}
\end{split}
\end{equation}

In the effective potential $U(a,\Phi)$ we have included the term proportional
to $P_\psi$ given that this momentum is conserved.

From the Hamiltonian constraint, Eq.~(\ref{E_campo_1}), and the Hamilton's equations we
obtain that the effective potential $U(a, \Phi)$ satisfies the following
conditions in order to have a static universe solution at $a=a_0$ and
$\Phi=\Phi_0$ in a JBD theory
\begin{eqnarray}
   U(a_0,\Phi_0)=0\,,\label{c1} \\
\nonumber \\
\frac{\partial U}{\partial a }(a_0,\Phi_0)=0\,,\label{c2}\\
\nonumber \\
\frac{\partial U}{\partial \Phi }(a_0,\Phi_0)=0\,.\label{c3}
\end{eqnarray}

The conditions (\ref{c1}, \ref{c2}, \ref{c3}) are satisfied if the following equations are fulfilled

\begin{equation}
  V_0' =  \frac{3}{a_0^2}\label{condicion_1}\,,
\end{equation}
\begin{equation}
P_\psi ^2 = 8 \pi ^4 a_0^4 \Phi _0 \label{condicion_2}\,,
\end{equation}
\begin{equation}
W_0 + V_0 = \frac{2 \Phi _0}{a_0^2}\,.\label{condicion_3}
\end{equation}

In order to study the stability of the static solution described
above, against small homogeneous and isotropic
perturbations, we study the Hamiltonian Eq.~(\ref{hamiltoniano_1}) near the
static solution. 
In order to do this, we consider small perturbations around the static solution
for the scale factor and the JBD field. We set
\begin{eqnarray}
a(t) &=& a_0 [1 + \varepsilon(t)], \\
\Phi(t) &=& \Phi_0[1+\beta(t)]  \label{condiciones_perturbacion},
\end{eqnarray}
where $\varepsilon\ll 1$ and $\beta\ll 1$ are small perturbations.

The Hamiltonian Eq.~(\ref{hamiltoniano_1}) near the static solution
is given by

\begin{eqnarray}\label{hamiltoniano_eta_beta}
\tilde{H}(\varepsilon,\beta) & =&-\frac{N \omega }{12 \pi^2 a_0^3
\Phi_0 (2 \omega +3)}\Big[P^2_{\varepsilon} -\frac{6 P_{\beta
}^2}{\omega } +\frac{6 P_{\beta } P_{\varepsilon}}{\omega }
-\frac{a_0^4\Phi_0^2 \left(144 \pi^4 (2 \omega +3)\right)
\varepsilon \,\beta}{\omega }\\\nonumber
\\\nonumber
&& -\frac{a_0^6 \Phi_0^3 \left(12 \pi^4 (2 \omega +3)\right)
V_0''\,\beta^2}{\omega}-\frac{a_0^4 \Phi _0^2 \left(288 \pi ^4 (2
\omega +3)\right) \varepsilon ^2}{\omega } \Big],
\end{eqnarray}
where $V_0'' = (d^2 V(\Phi)/ d\Phi^2)_{\Phi = \Phi_0}$. The conjugate momenta associated to $\varepsilon$ and $\beta$ are given by
\begin{equation}
P_\varepsilon = -\frac{6 \pi ^2 a_0^3 \Phi_0\,\dot{\beta}}{N}-\frac{12
\pi^2 a_0^3 \Phi_0\, \dot{\varepsilon}}{N},
\end{equation}
\begin{equation}
P_\beta = \frac{2 \pi ^2 a_0^3 \Phi_0 \omega\,\dot{\beta}
}{N}-\frac{6 \pi ^2 a_0^3 \Phi _0\,\dot{\varepsilon }}{N}.
\end{equation}

We can write Eq.~(\ref{hamiltoniano_eta_beta}) as follow
\begin{equation}
\tilde{H}=\frac{1}{2} Q_{ab}\, q^a q^b + \frac{1}{2} P^{ab}\, p_a
p_b, \;\;\;\; a,b = 1, 2 \label{hamiltoniano_tensorial}\,,
\end{equation}
where $q_a= (\varepsilon, \beta)$, $p_a= (P_\varepsilon, P_\beta)$.
The $2\times 2$ constant matrices $P$ and $Q$ are given by

\begin{equation}
P= -\frac{N}{2 \pi^2 a_0^3 \Phi_0 (2\omega +3)}\,
   \begin{pmatrix}
   \frac{\omega}{3}  & & 1  \\
   \\
   1 & & -2  \\
   \end{pmatrix},
\end{equation}
\vspace{0.1cm}
\begin{equation}
Q= N\pi ^2 a_0 \Phi_0\,\begin{pmatrix}
   48 & &12 \\
   \\
   12 & &2\,a_0^2 \Phi_0 V_0''\\
   \end{pmatrix}.
\end{equation}

By using the Hamilton equation in
Eq. (\ref{hamiltoniano_tensorial}) we obtain
\begin{equation}
\frac{d q^a}{d t} = \frac{\partial \tilde{H}}{\partial p_a} = P^{a
b} p_b\,\,.
\end{equation}

We take the derivative of this expression obtaining

\begin{equation}\label{oscilador}
\frac{d^2 q^a}{d t^2} = P^{a b} \frac{d p_b}{dt}= - P^{ab}\,
Q_{bc}\, q^c,
\end{equation}
where we have used that $\frac{d p_b}{dt}=- \frac{\partial
H}{\partial q^b}= -Q_{bc}\,q^c$. From Eq.~(\ref{oscilador}) we
notice that the eigenvalues of the matrix $\Lambda = P Q$ are the
square of the frequencies for the small oscillations around the
static solution. The eigenvalues of $\Lambda$ are
\begin{equation}\label{frecuencia_omega}
\lambda^2_{\pm} = \frac{N^2}{(3+2\omega)a^2_0}\left[a_0^2 \Phi_0
V_0 '' - 2(3+2\omega) \pm \sqrt{[a_0 ^2 \Phi_0 V_0 '' ]^2 +4a_0 ^2
\Phi_0 V_0 ''(3+2\omega) + 8\omega(3+2\omega) }\right].
\end{equation}
The static solution is stable if $\lambda_{\pm}^2 > 0$. Then
assuming that the parameter $\omega$ satisfies $(3+2\omega) > 0$, we
find that the following inequalities must be satisfied in order to
have a stable static solution

\begin{equation}
0 < a_{0}^2 \Phi_0 V_0'' <\frac{3}{2}\label{restriccion1}\,\,,
\end{equation}
\begin{equation}\label{restriccion2}
-\frac{3}{2} < \omega < -\frac{1}{4} \left[\sqrt{9 - 6 a_{0}^2
\Phi_0 V_0''} + (3+ a_{0}^2 \Phi_0 V_0'')\right].
\end{equation}

These inequalities restrict the parameters of the model, as it was discussed in Refs.~\cite{delCampo:2007mp, delCampo:2009kp}. The first imposes a condition on
the JBD potential, specifically for its first and second derivatives: $0 < V''_0 < V'_0/(2\Phi_0)$. The second inequality restricts the values of the JBD parameter.

We can note that, contrary to General Relativity, the static universe solution
could be stable against homogeneous and isotropic classical perturbations.

It is worth commenting on the regime of the JBD parameter selected by these conditions. The classical stability inequalities (\ref{restriccion1})-(\ref{restriccion2}) confine the JBD parameter to values for which the combination $2\omega+3$ lies in the small interval $0 < 2\omega+3 < 3/4$, approaching zero at the lower edge $\omega=-3/2$. Consequently, the proximity of the stable solutions to the boundary $2\omega+3=0$ is not a fine-tuned choice but a structural feature of the classical stability conditions.

Regarding the healthiness of the scalar sector, it is convenient to consider the Einstein frame, obtained through the conformal transformation $\tilde{g}_{\mu\nu} = \Phi\, g_{\mu\nu}$. In this frame the action (\ref{accion_JBD}) takes the form
\begin{equation}\label{Einstein_frame}
S_E=\int d^4x \sqrt{-\tilde{g}}\left[-\frac{1}{2}\tilde{\cal R} + \frac{2\omega+3}{4}\,\frac{\partial_\mu \Phi\, \partial^\mu \Phi}{\Phi^2} - \frac{V(\Phi)}{\Phi^2} + \dots\right],
\end{equation}
so that the whole kinetic term of the scalar is proportional to $(2\omega+3)$. Introducing the canonically normalized field $\tilde{\phi} = \sqrt{(2\omega+3)/2}\;\ln \Phi$, the scalar sector acquires the standard form $+\frac{1}{2}\partial_\mu\tilde{\phi}\,\partial^\mu \tilde{\phi}$. The ghost-free condition is therefore $2\omega + 3 > 0$, i.e.\ $\omega > -3/2$. This condition is automatically satisfied by the stable configurations of our model, since the classical stability inequalities enforce $\omega > -3/2$. As discussed above, however, these same conditions keep $2\omega+3$ close to zero, and the resulting smallness of the kinetic coefficient in the Einstein frame implies a strong coupling of the canonically normalized scalar to matter, which scales as $(2\omega+3)^{-1/2}$. This is precisely why the mass acquired by the JBD field through the potential $V(\Phi)$ plays an essential role in evading local-gravity constraints, as we discuss next.

Finally, we comment on the relation to observational constraints on Brans-Dicke theory. The stringent solar-system bound $\omega \gtrsim 4\times 10^{4}$~\cite{Bertotti:2003rm, Will:2014kxa} applies to a \emph{massless} Brans-Dicke field, whose long-range mediation modifies local gravity. In the present model the JBD field has a self-interacting potential $V(\Phi)$, which endows it with a mass and renders its mediation short-ranged, so that local-gravity tests do not directly constrain the value of $\omega$ in the regime considered here~\cite{Perivolaropoulos:2009ak}. We further emphasize that our analysis concerns the early-universe Einstein-static/emergent phase rather than the present-day local field; a complete assessment of the late-time viability of the model would require specifying the subsequent evolution of $\Phi$ toward its present value, which lies beyond the scope of the present work.

Having established the classical stability of the static solution and discussed the viability of the corresponding parameter regime, in the next section we turn our attention to its possible quantum instability, similar to the ones discussed in Refs.~\cite{Mithani:2011en, Mithani:2012ii, Mithani:2014jva, Mithani:2014toa, Mithani:2015ona}.

\section{Wheeler-DeWitt Equation}\label{sec:WdW}

In the context of quantum theory, the universe is described by
a wave function $\Psi(a, \Phi,\psi)$, and the conjugate momenta become
operators $P_a \to \hat{P}_a$, $P_\Phi \to \hat{P}_\Phi$, $P_\psi \to
\hat{P}_\psi$. In this context the Hamiltonian constraint
(\ref{constrainH}) is replaced by the Wheeler-DeWitt equation
$\mathcal{H} \Psi = 0$, see Ref.~\cite{Will}.

As discussed in Sec.~\ref{ES_JBD}, during the static regime the inflaton potential is flat and the momentum $P_\psi$ is conserved, so that $\psi$ is a cyclic variable. The full wave function can then be written in the separated form
\begin{equation}\label{ansatz_WDW}
\Psi(a,\Phi,\psi) = e^{\,i P_\psi \psi}\,\chi(a,\Phi),
\end{equation}
with $P_\psi$ a constant. Substituting this ansatz, the Wheeler-DeWitt equation reduces to an equation for the reduced wave function $\chi(a,\Phi)$,
\begin{eqnarray} \label{WDW_1}
\left[\hat{P}_a^2 - \frac{6\Phi^2 \hat{P}_\Phi^2}{\omega a} +
\frac{6\Phi \hat{P}_a \hat{P}_\Phi}{\omega a} +U(a,\Phi)\right]\chi(a,\Phi)=0,
\end{eqnarray}
where $P_\psi$ enters as a conserved parameter through the effective potential $U(a,\Phi)$ of Eq.~(\ref{Potencial(a,Ph)}). In what follows we work directly with the reduced wave function $\chi(a,\Phi)$.

Before proceeding, we note that the minisuperspace kinetic term, 
as can be read directly from the structure of the Hamiltonian in 
Eq.~(\ref{constrainH}), is not positive definite. This is a 
generic feature of canonical gravity in the ADM formulation: the 
gravitational degree of freedom associated with the scale factor 
contributes with a sign opposite to that of the matter sector, 
producing an indefinite metric on minisuperspace and rendering the 
Wheeler-DeWitt equation hyperbolic rather than elliptic. As a 
consequence, the standard mechanical interpretation of the sign of 
$U(a,\Phi)$ --- in terms of classically allowed and forbidden regions, valid for systems with positive-definite kinetic terms --- does not directly apply here. The physically meaningful structure is instead encoded in the full Hamiltonian constraint, $\mathcal{H} = K(a,\Phi;P_a,P_\Phi) + U(a,\Phi) = 0$, where $K$ denotes the kinetic term; this is a surface in the phase space $(a,\Phi;P_a,P_\Phi)$ rather than a curve in the $(a,\Phi)$ plane. The locus $U(a,\Phi) = 0$ marks where the kinetic term vanishes ($K=0$); it is contained in, but does not coincide with, the constraint $\mathcal{H}=K+U=0$, and contains in particular the zero-momentum configurations. 
The WKB estimates performed below should therefore be understood as 
quantifying the qualitative behavior of the effective potential along 
restricted directions in minisuperspace, with their physical 
interpretation made precise by the global analysis presented in 
Sec.~\ref{sec:ZERO_LOC}.

In order to obtain Eq.~(\ref{WDW_1}), we have used the minisuperspace approximation, which is
appropriated for our model where the universe is homogeneous,
isotropic and closed during the ES regime \cite{Vilenkin:1987kf}.
Also in equation (\ref{WDW_1}), we have not included the terms related to the
ambiguity in the ordering of the non-commuting factors in the
Hamiltonian. These terms do not affect the wave function in the
semiclassical regime and usually in the study of semiclassical
stability of ES state these terms are not included, see Refs.~\cite{Mithani:2011en,
Mithani:2012ii, Mithani:2014jva, Mithani:2014toa, Mithani:2015ona, delCampo:2015yfa}.

With the purpose of assessing the semiclassical stability of the ES universe against quantum tunneling processes, we analyze the properties of the effective WDW potential $U(a,\Phi)$, Eq.~\eqref{Potencial(a,Ph)}, in the neighborhood of the static configuration, which corresponds to a point satisfying $U=0$. In order to set the parameters of the model and of the JBD potential $V(\Phi)$, we take into account the classical stability conditions discussed in the previous section.

In the semiclassical WKB approximation, tunneling processes connecting configurations satisfying the Hamiltonian constraint can be formally characterized by an exponential suppression factor \cite{Mithani:2011en},
\begin{equation}
\mathcal{P}\sim e^{-2S},
\label{Tunel}
\end{equation}
where the WKB action is defined as
\begin{equation}
S = \int \sqrt{U(a,\Phi)} \, dl.
\label{Probabilidad}
\end{equation}

The differential $dl$ parametrizes possible paths in minisuperspace connecting the stable static configuration $(a_0,\Phi_0)$ to other configurations.

We stress that, in an indefinite minisuperspace metric, a fully consistent semiclassical treatment should be based on the Hamilton-Jacobi equation associated with the WDW operator,
\begin{equation}
G^{AB}\,\partial_A S\,\partial_B S + U(q) = 0,
\label{HJ}
\end{equation}
where $G^{AB}$ is the (Lorentzian) minisuperspace metric and $q^A=(a,\Phi)$. In this formulation the tunneling exponent is in general path dependent. The corresponding line-integral estimate, Eqs.~\eqref{Tunel}-\eqref{Probabilidad}, should therefore be regarded as heuristic, meaningful along the restricted trajectories considered below --- where one minisuperspace variable is held fixed and the induced kinetic term is positive definite --- but which does not by itself define a global tunneling amplitude. Accordingly, the calculations of this section are presented as illustrative diagnostics for specific restricted channels rather than as a determination of the global semiclassical stability of the static configuration.

Due to the indefinite signature of the minisuperspace kinetic term, the interpretation of the sign of $U(a,\Phi)$ must be understood in the context of the Hamiltonian constraint. In particular, the relevant endpoints of the tunneling channels considered here are zero-momentum configurations lying on the locus $U(a,\Phi)=0$; since the Einstein static state itself lies on $U=0$ with $P_a=P_\Phi=0$, these channels connect such configurations.

Then, in order to investigate the quantum instability discussed by Mithani and Vilenkin \cite{Mithani:2014jva, Mithani:2011en, Mithani:2012ii, Mithani:2014toa, Mithani:2015ona}, we focus on tunneling processes originating from the static configuration and leading to configurations with vanishing scale factor or vanishing JBD field. 
In this context, as a first illustrative approach, we consider trajectories along which the JBD field remains fixed at its equilibrium value ($\Phi=\Phi_0$), while the scale factor varies.

In this restricted setting, the WKB action in Eq.~\eqref{Probabilidad} reduces to
\begin{equation}
S = \int_{\epsilon}^{a_0} \sqrt{U(a,\Phi_0)} \, da,
\label{Probabilidad1}
\end{equation}
where $\epsilon$ denotes a regulator that will be taken to zero at the end of the calculation.

Near $a\approx 0$, the dominant contribution to the effective potential $U(a,\Phi)$ arises from the first term in Eq.~\eqref{Potencial(a,Ph)}, leading to
\begin{eqnarray}
S & \approx & \int_{\epsilon}^{a_0} \sqrt{-\frac{3 P_\psi^2 \Phi_0 (2 \omega +3)}{a^2 \omega }} \, da = \sqrt{\frac{-3\,P_{\psi}^2\,\Phi_0(2\omega +3)}{\omega}}\,\ln(a_0/\epsilon).
\end{eqnarray}

This yields an exponential suppression factor of the form
\begin{eqnarray}\label{condPa}
\mathcal{P}\sim \left(\frac{\epsilon}{a_0}\right)^{\sqrt{(-12\,P_{\psi}^2\,\Phi_0(2\omega +3))/\omega}}.
\end{eqnarray}

In the limit $\epsilon \rightarrow 0$, the suppression factor vanishes, indicating that tunneling processes along this restricted class of trajectories are strongly suppressed. This result is consistent with the interpretation that the ES universe is semiclassically stable against such tunneling channels.

On the other hand, one may consider tunneling processes in which the JBD field departs from its equilibrium value $\Phi_0$ while the scale factor remains fixed at $a=a_0$. In this case, the WKB action is given by
\begin{equation}
S = \int_{\epsilon}^{\Phi_0} \sqrt{U(a_0,\Phi)} \, d\Phi.
\label{Probabilidad2}
\end{equation}

Near $\Phi\approx 0$, the dominant contribution to the effective potential $U(a,\Phi)$ arises from the fourth term in Eq.~\eqref{Potencial(a,Ph)}, leading to
\begin{eqnarray}
S &=& \int_{\epsilon}^{\Phi_0} \sqrt{-\frac{24 \pi ^4 a_{0}^4 \Phi (2 \omega +3) V(\Phi )}{\omega }} \, d\Phi.
\end{eqnarray}

Assuming that the JBD potential behaves near $\Phi\sim 0$ as $V(\Phi) \sim C/\Phi^{\alpha}$, with $C$ and $\alpha$ positive constants, one finds
\begin{eqnarray}
S &=& \left(\frac{-2}{\alpha -3}\right)\sqrt{-\frac{24 \pi ^4 a_{0}^4 (2 \omega +3) C}{\omega }}\left[\Phi^{(3-\alpha)/2}_0 - \epsilon^{(3-\alpha)/2}\right].
\end{eqnarray}

The associated semiclassical suppression factor then takes the form
\begin{eqnarray}
\mathcal{P} &\sim & \exp\left\{-\frac{D}{\epsilon^{\frac{(\alpha - 3)}{2}}}\right\},
\end{eqnarray}
where $D = \frac{8 \pi^2}{(\alpha - 3)}\sqrt{-\frac{6 C a_0^4 (3 + 2\omega)}{\omega}}$. In the limit $\epsilon\rightarrow 0$, this suppression factor vanishes for $\alpha > 3$, indicating that tunneling processes along this restricted class of trajectories are strongly suppressed.

Notice that, in the restricted classes of trajectories considered above, the kinetic term effectively becomes positive definite along the path. In this situation, configurations with $U(a,\Phi)>0$ can be interpreted as defining a potential barrier, in analogy with standard quantum mechanical tunneling. 

Taken together, the above results indicate that the tunneling channels considered here --- variations of the scale factor at fixed $\Phi$, and of the JBD field at fixed $a$ --- are strongly suppressed, provided the JBD potential satisfies suitable conditions near $\Phi=0$. These findings complement previous analyses of classical stability \cite{delCampo:2007mp, delCampo:2009kp} and suggest that, at least along these channels, the Mithani--Vilenkin instability can be avoided in certain regions of parameter space within JBD theory. They do not, however, by themselves establish the global semiclassical stability of the configuration, which would require the Hamilton--Jacobi analysis mentioned above, including a full exploration of tunneling paths connecting different $U(a,\Phi)=0$ configurations in minisuperspace; this is left for future work.

It is important to note that the WKB estimates presented above are based on restricted trajectories in minisuperspace and do not, in general, correspond to complete tunneling paths connecting configurations that satisfy the Hamiltonian constraint. In particular, the endpoints of these trajectories are not necessarily located on the surface $U(a,\Phi)=0$. Therefore, these calculations should be understood as providing indications of possible tunneling directions, while the identification of physically admissible initial and final states requires a global analysis of the zero loci of the WDW potential.

\begin{figure}[t]
\centering
\subfigure[ ]{\includegraphics[scale=0.54]{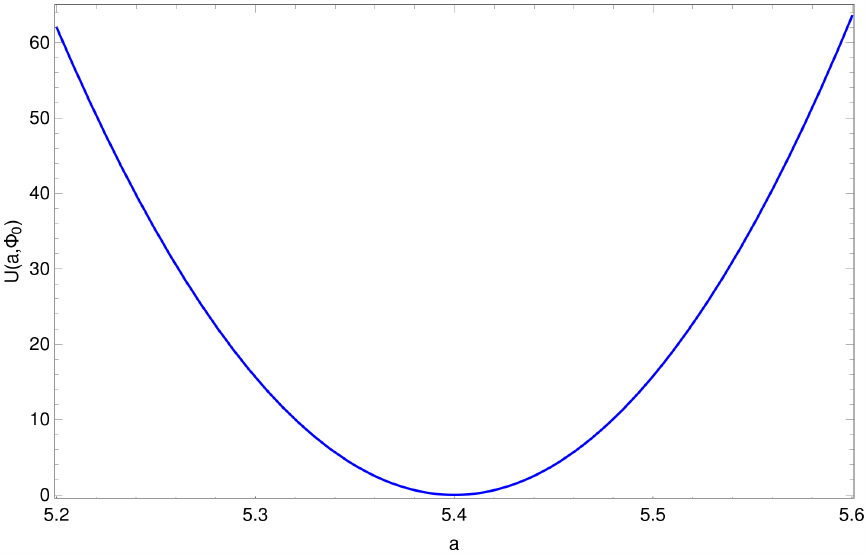}}
\subfigure[]{\includegraphics[scale=0.56]{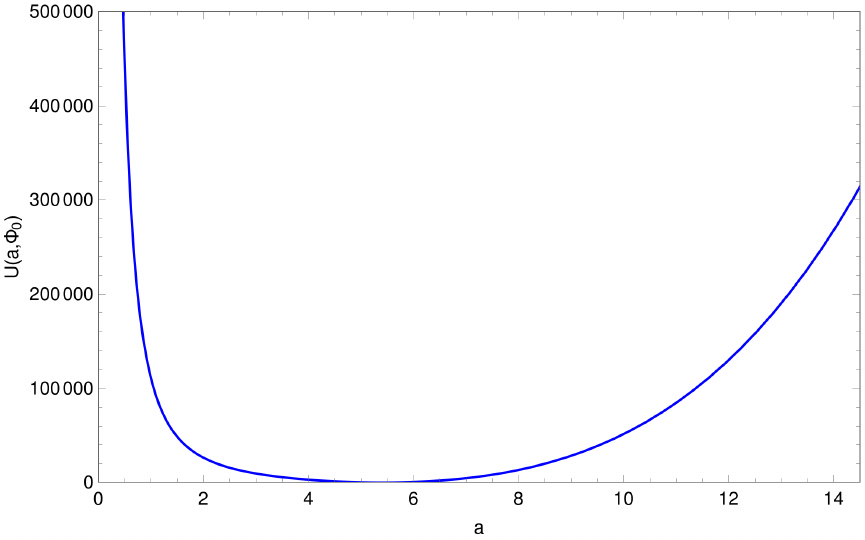}}
\caption{The plots were obtained by using the JBD  potential, Eq. \eqref{Potencial_BD_1}, and by considering $a_0=5.4$, $\Phi_0=0.9$ and $\omega=-1.45$. In the plots (a) and (b) we assume that the value of the $\Phi$ field is fixed and the scale factor $a$ is varied.}
\label{fig:evolucion1-1}
\end{figure}

\section{SPECIFIC EXAMPLES}\label{sec:Examples}

As a first example we will consider the model studied in Refs.~\cite{delCampo:2007mp, delCampo:2009kp}. In these works, the parameters of the JBD model were adjusted to satisfy the classical equilibrium conditions of the static universe solution at $a=a_0$ and $\Phi= \Phi_0$.
Particularly, these works consider a JBD potential similar to the following
\begin{eqnarray}
    V(\Phi )= V_0 + A \left(\Phi -\Phi _0\right)+\frac{1}{2} B \left(\Phi -\Phi _0\right)^2\label{Potencial_BD_1}.
   \end{eqnarray}
The constants in the potential (\ref{Potencial_BD_1}) were fixed by considering the classical stability conditions of the ES solution discussed in the previous section. Then we have
   
\begin{equation}\label{parametro1}
V_0 = \frac{2\Phi_0}{a_0^2} - W_0,
\end{equation}
\begin{equation}\label{parametro2}
A = \frac{3}{a_0^2},
\end{equation}
\begin{equation}\label{parametro3}
B = \frac{X}{a_0 ^2  \Phi_0}.
\end{equation}
The dimensionless parameter $X$, satisfy $0 < X < 3/2$ and the JBD parameter requires
\begin{equation}
-\frac{3}{2} < \omega < - \frac{\sqrt{3}}{4} \sqrt{3 - 2 X} - \frac{1}{4} (3 + X).
\end{equation}

In order to plot the effective potential $U(a, \Phi)$ we take the following values for the parameters in the JBD potential $\Phi_0 = 0\ldotp9$, $a_0 = 5\ldotp4$, $X = 1$ and $\omega = -1\ldotp45$, where units are such that $8 \pi G = 1$ and $c = \hbar = 1$. These particular parameters satisfy all the classical equilibrium conditions of the static Universe solution discussed previously, for the static Universe solution at $a=a_0$ and $\Phi= \Phi_0$.

 \begin{figure}[t]
    \centering
    \subfigure[ ]{\includegraphics[scale=0.54]{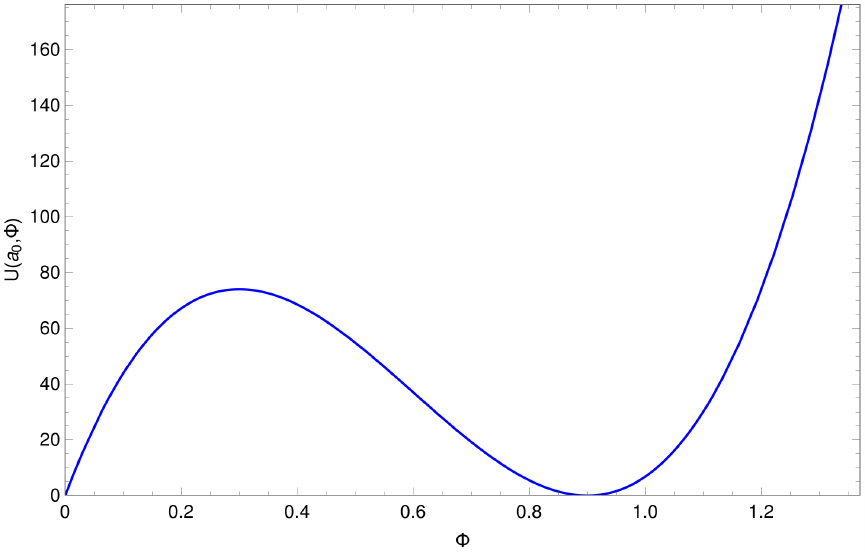}}
    \subfigure[]{\includegraphics[scale=0.56]{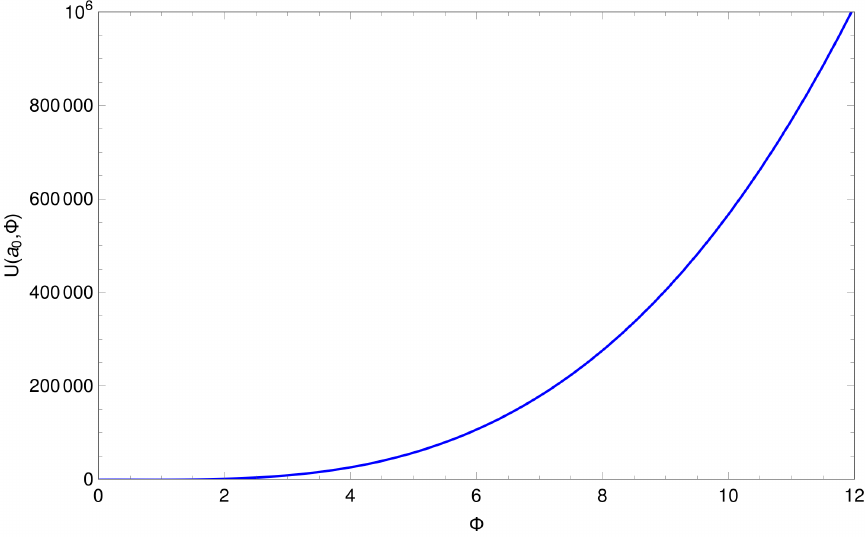}}
    \caption{The plots were obtained by using the JBD potential, Eq. \eqref{Potencial_BD_1}, and by considering $a_0=5.4$, $\Phi_0=0.9$ and $\omega=-1.45$. In figures (a) and (b), we consider the case where the value of the scale factor remains constant at $a = a_0$, while the JBD field $\Phi$ is varied.}
    \label{fig:evolucion1.2}
   \end{figure}

In Figs.~(\ref{fig:evolucion1-1}) and (\ref{fig:evolucion1.2}) we 
plot the effective potential $U(a,\Phi)$ corresponding to the JBD 
potential~(\ref{Potencial_BD_1}) and the parameters discussed above. 
In particular, in Fig.~(\ref{fig:evolucion1-1}) we consider 
$\Phi = \Phi_0$ fixed and plot $U(a, \Phi_0)$ as a function of the 
scale factor $a$. The effective potential exhibits a local minimum 
at $a_0 = 5.4$, which is classically stable. Panel (a) shows the 
potential near the equilibrium point $a \sim a_0$, while panel (b) 
displays a wider region of $a$. Since $U(a, \Phi_0)$ diverges as 
$a \to 0$, the WKB action along this direction in minisuperspace 
diverges as well, strongly suppressing tunneling along this 
trajectory, in accordance with the analysis of Sec.~\ref{sec:WdW}.

In Fig.~(\ref{fig:evolucion1.2}) we consider $a = a_0$ fixed and 
plot $U(a_0, \Phi)$ as a function of the JBD field $\Phi$. The 
effective potential exhibits a local minimum at $\Phi = \Phi_0 = 0.9$, 
which is classically stable. However, from Fig.~(\ref{fig:evolucion1.2})-(a), 
it can be observed that a finite-sized barrier separates $\Phi_0$ 
from $\Phi = 0$. Consequently, the static universe solution discussed 
in Ref.~[91], although classically stable, admits a semiclassical 
tunneling channel toward $\Phi = 0$ within the WKB approximation 
considered. This situation can be remedied by including a term in 
the JBD potential that corrects its behavior as the JBD field 
approaches zero, in accordance with the discussion of the previous 
section.

\begin{figure}[t]
     \centering
    \subfigure[ ]{\includegraphics[scale=0.54]{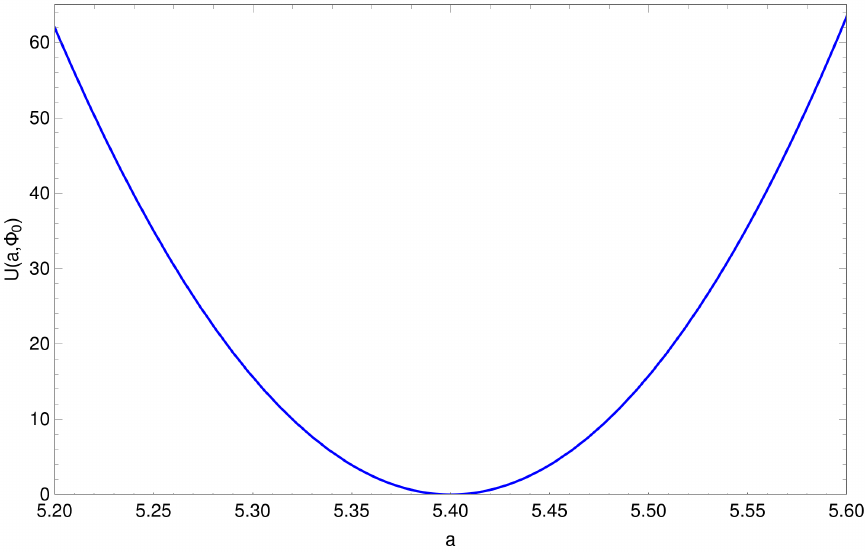}}
    \subfigure[]{\includegraphics[scale=0.56]{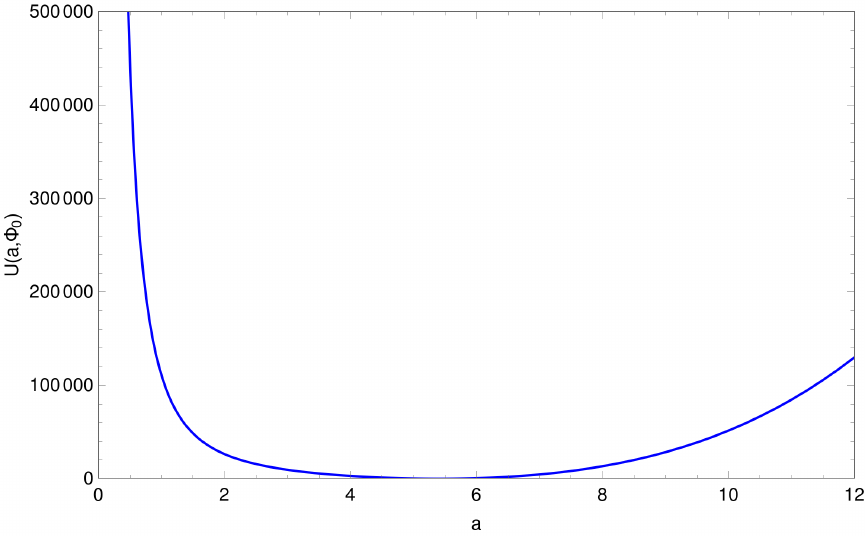}}
    \caption{ The plots were obtained by using the Brans-Dicke potential \eqref{Potencial_BD_2}, and by considering $a_0=5.4$, $\Phi_0=0.9$ and $\omega=-1.45$. In the plots (a) and (b) we assume that the value of the Φ field is fixed at $\Phi=\Phi_0$ and the scale factor $a$ is varied.}
        \label{fig:evolucion1.3}
   \end{figure}

As a second example, we consider a model similar to the one developed in Ref.~\cite{delCampo:2007mp}, but where the JBD potential has been corrected to incorporate the term discussed in the previous section, which suppresses tunneling toward $\Phi  \to 0$.

\begin{eqnarray}
   V(\Phi)= A \left(\Phi -\Phi _0\right)+\frac{1}{2} B \left(\Phi -\Phi _0\right){}^2+\frac{C}{\Phi ^6}\,,
   \label{Potencial_BD_2}
   \end{eqnarray}
this potential takes into account the classical stability conditions discussed in Section \ref{ES_JBD}, but also, the stability conditions under quantum tunneling discussed in the previous Section. 

We stress that the inverse-power contribution $C/\Phi^6$ in Eq.~(\ref{Potencial_BD_2}) is introduced as a phenomenological barrier whose role is to suppress the evolution of the system toward $\Phi \to 0$. Its physical motivation rests on the observation that, since the effective gravitational coupling scales as $G_{\rm eff} \propto 1/\Phi$, the limit $\Phi \to 0$ corresponds to the strong-coupling regime in which the scalar-tensor description breaks down. A potential that diverges as $\Phi \to 0$ therefore encodes, in a simple and economical way, the expectation that this regime is dynamically inaccessible. We further emphasize that our conclusions do not depend on the specific exponent: as shown in Sec.~\ref{sec:ZERO_LOC}, only the asymptotic behavior $V(\Phi) \sim C/\Phi^\alpha$ with $\alpha > 3$ near $\Phi = 0$ is required, the value $\alpha = 6$ being chosen merely for definiteness. Inverse-power potentials of this kind arise naturally in scalar-field cosmology and in nonperturbative constructions within supergravity and string-inspired scenarios~\cite{Ratra:1987rm, Binetruy:1998rz}, although we do not attempt to derive the present barrier from any specific ultraviolet completion; a first-principles justification lies beyond the scope of this work.

The parameters $V_0$, $A$, $B$ are given by
\begin{equation}\label{parametro2.1}
   V_0 = \frac{C}{\Phi_0^6}\,,
   \end{equation}
    \begin{equation}\label{parametro2.2}
    A = \frac{3}{a_0 ^2} + \frac{6 C}{\Phi_0 ^7} \,,
   \end{equation}
    \begin{equation}\label{parametro2.3}
    B = \frac{X}{a_0^2 \Phi_0} - \frac{42 C}{\Phi_0 ^8}\,.
   \end{equation}

\begin{figure}[t]
    \centering
    \subfigure[ ]{\includegraphics[scale=0.54]{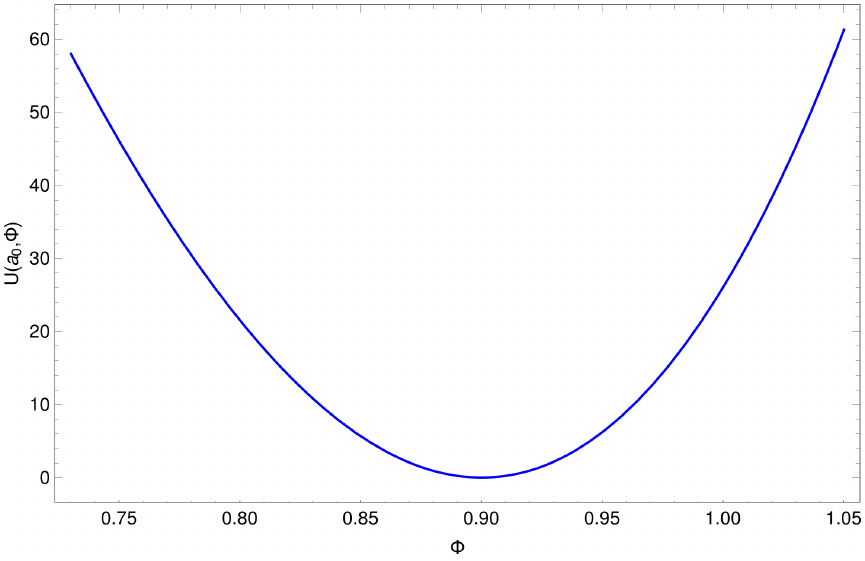}}
    \subfigure[]{\includegraphics[scale=0.56]{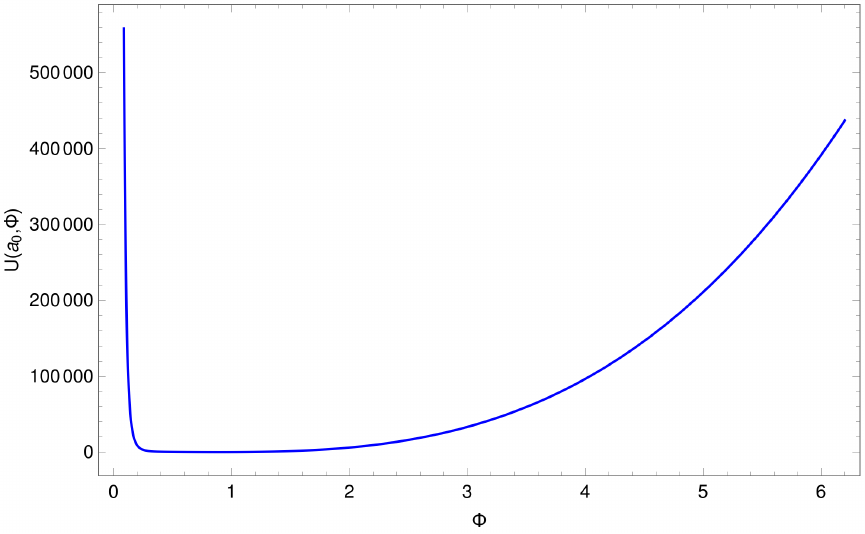}}
    \caption{The plots were obtained by using the JBD potential \eqref{Potencial_BD_2}, considering $a_0=5.4$, $\Phi_0=0.9$ and $\omega=-1.45$. In figures (a) and (b), we consider the case where the value of the scale factor remains constant at $a = a_0$, while the JBD field $\Phi$ is subjected to variations.}
    \label{fig:evolucion1.4}
   \end{figure}

Similar to the previous example, the dimensionless parameter $X$ satisfy $0 < X < 3/2$ and  the JBD parameter requires $-\frac{3}{2} < \omega < - \frac{\sqrt{3}}{4} \sqrt{3 - 2 X} - \frac{1}{4} (3 + X)$.
For this case we consider $\Phi_0 = 0\ldotp9$, $a_0 = 5\ldotp4$, $C = 0\ldotp00002$,  $X = 1$ and $\omega = -1\ldotp45$, these numerical values satisfy the classical stability conditions, Eqs. (\ref{restriccion1})-(\ref{restriccion2}) mentioned in section III. Units are such that $8 \pi G = 1$ and $c = \hbar = 1$.
The effective potential $U(a, \Phi)$ is plotted in Figs.~(\ref{fig:evolucion1.3}, \ref{fig:evolucion1.4}), taking into account the JBD potential (\ref{Potencial_BD_2}) and the parameters discussed above.

In Fig.~(\ref{fig:evolucion1.3}) we consider $\Phi = \Phi_0$ fixed 
and plot $U(a, \Phi_0)$ as a function of the scale factor $a$. As 
in the first example, the effective potential exhibits a local 
minimum at $a_0 = 5.4$, which is classically stable. Since 
$U(a, \Phi_0)$ diverges as $a \to 0$, the WKB action along this 
direction in minisuperspace diverges, strongly suppressing tunneling 
along this trajectory, consistent with the analysis of the previous section.

In Fig.~(\ref{fig:evolucion1.4}) we consider $a = a_0$ fixed and 
plot $U(a_0, \Phi)$ as a function of the JBD field $\Phi$. The 
effective potential exhibits a local minimum at $\Phi_0 = 0.9$, 
which is classically stable. Furthermore, since $U(a_0,\Phi)$ 
diverges as $\Phi \to 0$, the WKB action along this direction in 
minisuperspace diverges as well, strongly suppressing tunneling 
along this trajectory, consistent with the analysis of 
Sec.~\ref{sec:WdW}.

Therefore, we can conclude that within the context of a JBD theory, a static universe solution can be achieved. This solution remains stable under classical perturbations, similar to those examined in previous studies \cite{delCampo:2007mp, delCampo:2009kp}. 
Also, the present analysis indicates that the solution is not generically subject to the quantum instability identified by Mithani and Vilenkin \cite{Mithani:2014jva, Mithani:2011en, Mithani:2012ii, Mithani:2014toa, Mithani:2015ona}, within the class of tunneling channels considered.

\section{Zero loci of the Wheeler-DeWitt potential and tunneling endpoints}
\label{sec:ZERO_LOC}

In order to extend the analysis presented in the previous sections, it is necessary to consider the global structure of the Wheeler-DeWitt potential in minisuperspace. In particular, it is useful to analyze the zero loci of the effective potential $U(a,\Phi)$, which, as discussed in Sec.~\ref{sec:WdW}, mark the configurations of vanishing kinetic energy ($K=0$).
These loci carry useful information about the qualitative semiclassical structure, since the endpoints of the tunneling channels considered in Sec.~\ref{sec:WdW} are zero-momentum configurations lying on them.

It is worth noting that the zero loci of $U(a,\Phi)$ exhibit a nontrivial structure in minisuperspace, forming continuous curves rather than isolated configurations. In particular, the classically stable Einstein static solution $(a_0,\Phi_0)$ lies at the intersection of these curves. 
The structure of the zero loci of $U(a,\Phi)$ plays a central role in identifying the semiclassical tunneling endpoints that are physically accessible from the static configuration.

\begin{figure}[t]
\centering
\subfigure[ ]{\includegraphics[scale=0.33]{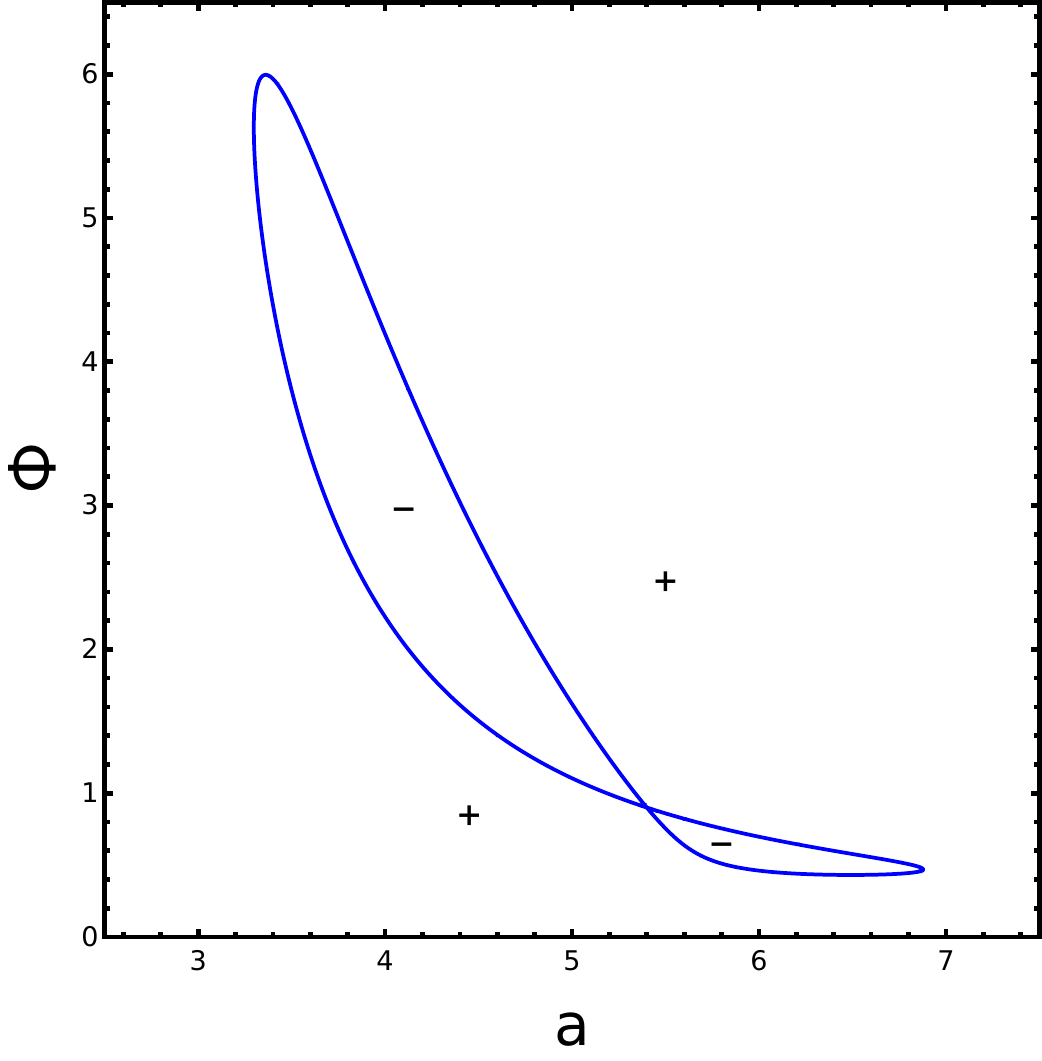}}
\subfigure[]{\includegraphics[scale=0.35]{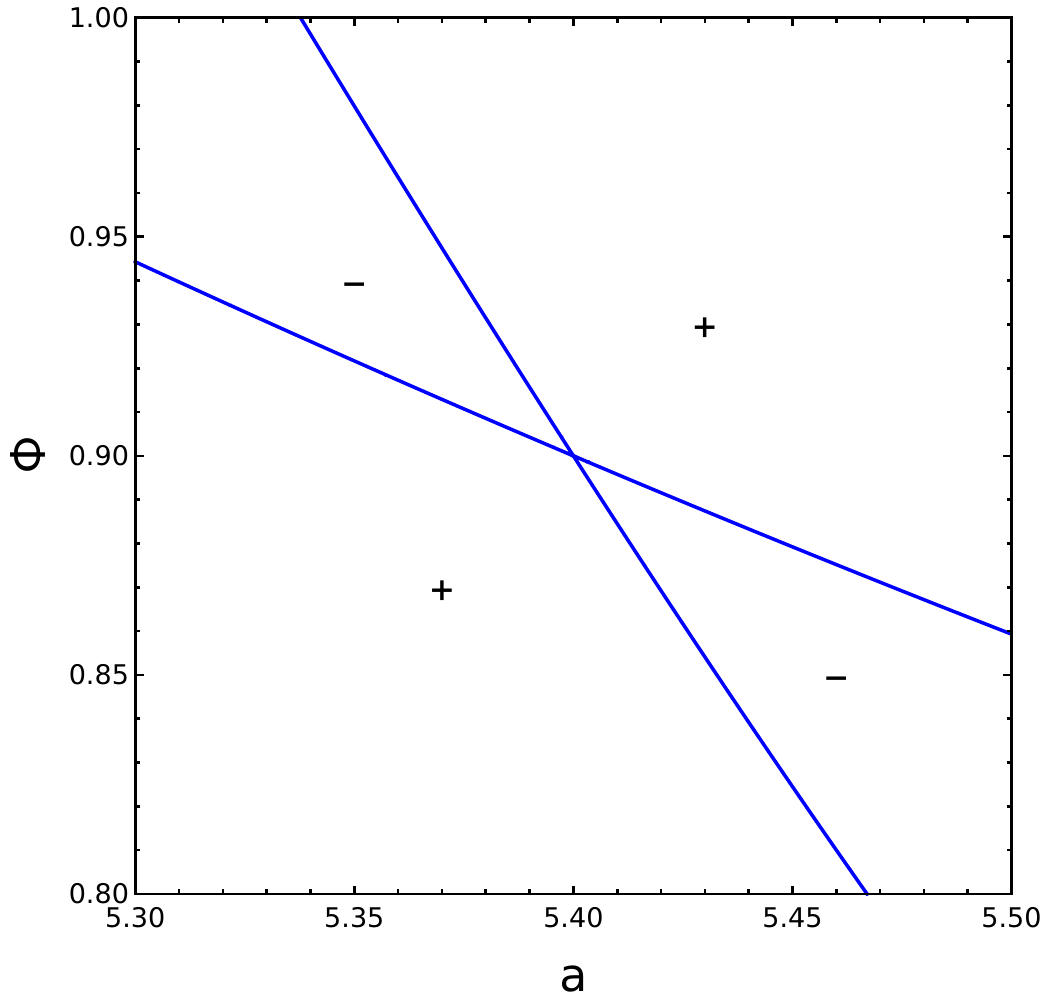}}
\caption{Zero loci of the effective Wheeler-DeWitt potential $U(a,\Phi)$ in minisuperspace. (a) Global structure of the curve $U(a,\Phi)=0$. (b) Zoomed-in view of the region surrounding the Einstein static solution, located at the intersection of the zero-potential curves. The signs $(+/-)$ indicate the regions where $U(a,\Phi)$ is positive and negative, respectively. Due to the indefinite signature of the minisuperspace kinetic term, the sign of $U$ does not admit a purely mechanical interpretation. In these figures we have considered the JBD potential (\ref{Potencial_BD_2}) and the parameter values of the second example in Sec.~\ref{sec:Examples}.}
\label{fig:Cont1}
\end{figure}

Figure~\ref{fig:Cont1} displays the curves defined by $U(a,\Phi)=0$ for representative choices of the model parameters consistent with classical stability, and for a JBD potential satisfying the condition $V(\Phi) \sim C/\Phi^\alpha$ with $\alpha > 3$ near $\Phi = 0$, discussed in Sec.~\ref{sec:WdW}.
These curves intersect at the Einstein static configuration and separate regions of minisuperspace with different semiclassical character. Due to the indefinite signature of the minisuperspace kinetic term, the interpretation of the sign of $U(a,\Phi)$ does not follow the standard intuition valid for positive-definite systems. 
In particular, regions where the effective potential acts as a barrier correspond to configurations that cannot be connected by classical trajectories satisfying the Hamiltonian constraint, whereas regions where $U(a,\Phi)$ effectively allows classical motion correspond to classically permitted domains.

A central observation that follows from the structure of $U(a,\Phi)$ concerns the singular configurations toward which the Mithani-Vilenkin instability would drive the system, namely, those with vanishing scale factor or vanishing JBD field. Inspection of the effective potential, Eq.~(\ref{Potencial(a,Ph)}), shows that the term proportional to $P_\psi^2$ scales as $1/a^2$ near $a = 0$ and is positive for parameters satisfying the classical stability conditions, ensuring that $U(a \to 0,\Phi) \to +\infty$ for any $\Phi > 0$. This conclusion is independent of the specific form of the JBD potential $V(\Phi)$ and holds in both examples considered in Sec.~\ref{sec:Examples}. 
Likewise, for the class of JBD potentials with $V(\Phi) \sim C/\Phi^\alpha$ and $\alpha > 3$ introduced in our second example, the contribution of 
the scalar potential to $U(a,\Phi)$ diverges as $\Phi \to 0$, so that $U(a,\Phi \to 0) \to +\infty$ as well. Consequently, the singular configurations do not lie on the locus $U(a,\Phi) = 0$. Since the tunneling channels considered here originate from the static state and connect zero-momentum configurations ($P_a=P_\Phi=0$), whose turning points necessarily satisfy $U=0$, the singular configurations cannot serve as endpoints of such channels, consistent with and more general than the local WKB estimates of Sec.~\ref{sec:WdW}. We stress, however, that this applies to the channels considered here and does not exclude $a \to 0$ or $\Phi \to 0$ as endpoints of more general semiclassical trajectories; a definitive statement would require the global Hamilton-Jacobi analysis mentioned in Sec.~\ref{sec:WdW}, which we leave for future work.


\begin{figure}[t]
\centering
\subfigure{\includegraphics[scale=0.55]{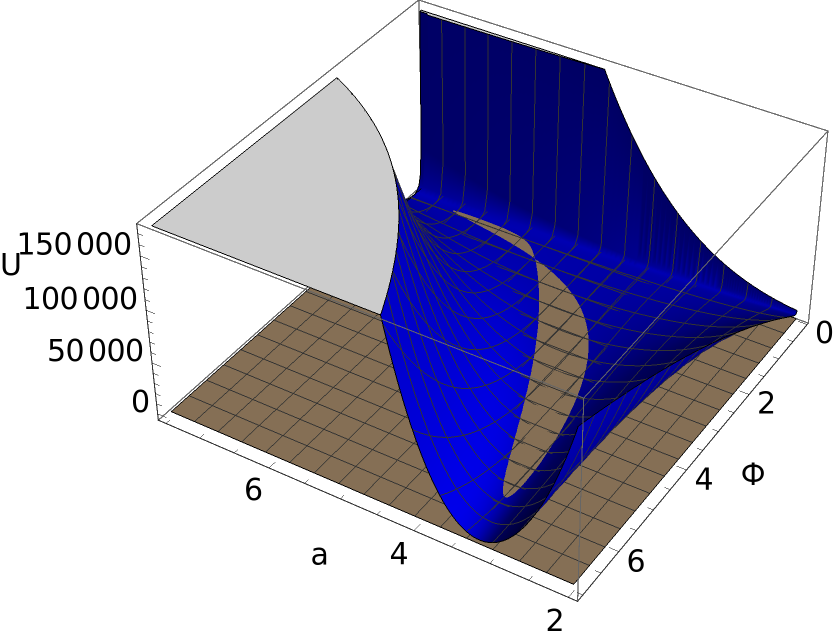}}
\caption{Three-dimensional representation of the effective Wheeler--DeWitt potential $U(a,\Phi)$ together with the plane $U=0$. 
In this plot we have considered the JBD potential (\ref{Potencial_BD_2}) and the parameter values of the second example in Sec.~\ref{sec:Examples}.}
\label{fig:3D}
\end{figure}

The zero loci shown in the figures thus define a restricted set of admissible tunneling channels connecting the ES universe to other configurations in minisuperspace. Importantly, these configurations are not limited to the singular limit $a\to0$, but instead form continuous curves determined by the structure of the WDW potential. This observation clarifies the role of the WKB estimates performed in the previous subsection: the explicit calculations correspond to particular restricted trajectories in minisuperspace, while the endpoints of the channels considered here are constrained to lie on the $U(a,\Phi)=0$ locus.

While a complete determination of the dominant semiclassical tunneling trajectory connecting different points on the $U(a,\Phi)=0$ locus would require a full variational analysis in minisuperspace, the qualitative structure revealed by the zero loci already provides important physical insight. In particular, it shows that the existence and suppression of tunneling channels are controlled not only by local properties of the potential near the static solution, but also by its global structure in minisuperspace.

Figure~\ref{fig:Cont1}\,(a) displays the global structure of the zero loci of the effective WDW potential $U(a,\Phi)$ in minisuperspace. 
Figure~\ref{fig:Cont1}~(b) shows a zoomed-in view of the region surrounding the ES universe, which corresponds to the intersection point of the zero-potential curves. 
Figure~\ref{fig:3D} presents a three-dimensional visualization of the effective WDW potential $U(a,\Phi)$, together with the plane $U=0$.

This analysis supports the interpretation that, within suitable regions of parameter space, the Einstein static universe in a JBD theory can be semiclassically robust against quantum decay to vanishing scale factor, while leaving open the possibility of more general tunneling processes whose detailed investigation is deferred to future work.

\section{Conclusions}\label{sec:conclu}

In this work we have investigated the classical and semiclassical stability of the Einstein static universe within the framework of Jordan-Brans-Dicke theory, in the context of the Emergent Universe scenario. We have shown that, for suitable choices of the JBD potential and model parameters, the Einstein static configuration can be classically stable, and the quantum instability leading to tunneling toward a vanishing scale factor is not generic and can be avoided.

Throughout, we have distinguished three logically different notions of stability, which should not be conflated: (i) classical perturbative stability, i.e.\ stability of the static solution under homogeneous and isotropic perturbations, established in Sec.~\ref{ES_JBD}; (ii) semiclassical stability against the specific, restricted tunneling channels analyzed here within the WKB approximation; and (iii) full quantum (semiclassical) stability, which would require a global analysis of all admissible tunneling trajectories.

Using the Wheeler-DeWitt equation in minisuperspace, we analyzed the structure of the effective potential governing the quantum dynamics of the system. Particular attention was devoted to the zero loci of the WDW potential, which mark the configurations of vanishing kinetic energy ($K=0$). We showed that the Einstein static universe lies at the intersection of these loci and that their geometry plays a crucial role in determining the available semiclassical tunneling channels.

Representative WKB estimates were performed for restricted classes of tunneling trajectories in minisuperspace, corresponding to variations of the scale factor or the JBD field while the remaining variable is held fixed. These calculations indicate that tunneling channels associated with configurations approaching a vanishing scale factor or vanishing JBD field can be strongly suppressed for a wide class of JBD potentials, including inverse power-law behaviors near the origin ($\Phi\sim 0$).
Furthermore, the analysis of the zero loci of the effective Wheeler-DeWitt potential, carried out in Sec.~\ref{sec:ZERO_LOC}, 
provides a structural argument independent of the specific WKB path: 
since $U(a,\Phi)$ diverges as $a \to 0$ and as $\Phi \to 0$, the singular 
configurations cannot act as turning points of the tunneling channels 
originating from the Einstein static state, which are therefore strongly 
suppressed. This reinforces the local WKB estimates; it does not, however, rule out more general semiclassical trajectories, whose analysis requires the 
Hamiltonian constraint $\mathcal{H}=K+U=0$ and is left for future work.

Then, our results suggest that the quantum instability leading to tunneling toward configurations of vanishing radius, previously identified in other ES constructions, is not universal and may be avoided in JBD realizations of the EU scenario. While a complete assessment of global semiclassical stability would require the identification of the dominant tunneling trajectories connecting different zero-potential configurations, the present analysis demonstrates that the structure of the Wheeler-DeWitt potential already imposes strong constraints on possible decay channels.

These findings provide further support for the viability of Emergent Universe models based on scalar-tensor theories and motivate future investigations of more general tunneling paths and their implications for the early-universe dynamics. Finally, since the analysis was carried out in the closed ($k=+1$) sector, our results complement recent work on geodesically complete FRW cosmologies~\cite{Kinney:2026abc, Easson:2026abc, Burwig:2026abc, Burwig:2025lyq} by addressing the semiclassical stability of such solutions, an aspect not captured by purely classical completeness criteria.

\section{Acknowledgements}
 P. L. was partially supported by Direcci\'on de Investigaci\'on y Creaci\'on Art\'{\i}stica de la Universidad del B\'{\i}o-B\'{\i}o through Grants FAPEI 2220228-FP, RE2320212 and GI2310339.
J. O. was supported by Direcci\'on de Postgrado  Universidad del B\'{\i}o-B\'{\i}o and by Research Assistant Grant of Escuela de Graduados Universidad del B\'{\i}o-B\'{\i}o. J. O. acknowledges the support of the ANID doctoral fellowship 21220369 and to the Systems Engineering Program at  Universidad Cooperativa de Colombia for continuous support and encouragement.


\begin{thebibliography}{99}

\bibitem{Mithani:2011en}
A.~T.~Mithani and A.~Vilenkin,
JCAP \textbf{01} (2012), 028
doi:10.1088/1475-7516/2012/01/028
[arXiv:1110.4096 [hep-th]].

\bibitem{Mithani:2012ii}
A.~Mithani and A.~Vilenkin,
[arXiv:1204.4658 [hep-th]].

\bibitem{Mithani:2014jva}
A.~T.~Mithani and A.~Vilenkin,
JCAP \textbf{05} (2014), 006
doi:10.1088/1475-7516/2014/05/006
[arXiv:1403.0818 [hep-th]].

\bibitem{Mithani:2014toa}
A.~T.~Mithani and A.~Vilenkin,
JCAP \textbf{07} (2015), 010
doi:10.1088/1475-7516/2015/07/010
[arXiv:1407.5361 [hep-th]].

\bibitem{Mithani:2015ona}
  A.~T.~Mithani and A.~Vilenkin,
  Phys.\ Rev.\ D {\bf 91}, no. 12, 123511 (2015)
  [arXiv:1503.00400 [hep-th]].

\bibitem{weinberg} S. Weinberg, Gravitation and Cosmology: Principle and Application of the General Relativity, Wiley, NY, 1972; Ch. W. Misner, K. S. Turner, J. A. Wheeler, Gravitation, W. H.: Freeman and Company, SF 1973.
\bibitem{peebles} P. J. E. Peebles, Principles of Physical Cosmology, Princeton University Press 1993; J. A. Peacock, Cosmological Physics, Cambridge University Press, 1998); S. Weinberg, Cosmology, Oxford University Press, USA, 2008.
\bibitem{kolb} E. Kolb and M. Turner, The Early Universe, Addison- Wesley Publishing(1989).

\bibitem{Guth1}  Guth A.,  The inflationary universe:
A possible solution to the horizon and flatness problems, 1981 Phys.
Rev. D {\bf 23} 347.

\bibitem{Albrecht}  Albrecht A. and  Steinhardt P. J.,
Cosmology for grand unified theories with radiatively induced symmetry
breaking, 1982 Phys. Rev. Lett. {\bf 48} 1220.

\bibitem{Linde1}  Linde A. D.,  A new inflationary universe scenario: A possible solution of the horizon,
flatness, homogeneity, isotropy and primordial monopole problems,
1982 Phys. Lett. {\bf 108B} 389.

\bibitem{Linde2}  Linde A. D.,  Chaotic inflation, 1983 Phys. Lett. {\bf 129B} 177.



\bibitem{Borde:1993xh}
  Borde A. and Vilenkin  A.,
  Eternal inflation and the initial singularity,
 1994 Phys.\ Rev.\ Lett.\  {\bf 72} 3305.

\bibitem{Borde:1997pp}
  Borde A. and Vilenkin A.,
  Violation of the weak energy condition in inflating spacetimes, 1997
  Phys.\ Rev.\  D {\bf 56} 717.

\bibitem{Guth:1999rh}
  A.~H.~Guth,
  ``Eternal inflation,''
  Annals N.\ Y.\ Acad.\ Sci.\  {\bf 950}, 66 (2001)
  [astro-ph/0101507].


\bibitem{Borde:2001nh}
  ~Borde A., ~Guth A.~H. and ~Vilenkin A.,
  Inflationary space-times are incompletein past directions, 2003
  Phys.\ Rev.\ Lett.\  {\bf 90} 151301.


\bibitem{Vilenkin:2002ev}
  ~Vilenkin A.,
  Quantum cosmology and eternal inflation,
  arXiv:gr-qc/0204061.



\bibitem{Ellis:2002we}
G.~F.~R.~Ellis and R.~Maartens,
Class. Quant. Grav. \textbf{21} (2004), 223-232
doi:10.1088/0264-9381/21/1/015
[arXiv:gr-qc/0211082 [gr-qc]].

\bibitem{Ellis:2003qz}
G.~F.~R.~Ellis, J.~Murugan and C.~G.~Tsagas,
Class. Quant. Grav. \textbf{21} (2004) no.1, 233-250
doi:10.1088/0264-9381/21/1/016
[arXiv:gr-qc/0307112 [gr-qc]].


\bibitem{Eddington} A. S. Eddington, Mon. Not. Roy. Astron. Soc. 90, 668-768 (1930); A. S. Eddington, Mon. Not. Roy. Astron. Soc. 92, 3-6 (1931).

\bibitem{Harrison:1967zz}
E.~R.~Harrison,
Rev.\ Mod.\ Phys.\  {\bf 39}, 862 (1967).

\bibitem{Gibbons:1987jt}
~Gibbons G.~W.,
The entropy and stability of the universe, 1987
Nucl.\ Phys.\  B {\bf 292} 784 .

\bibitem{Gibbons:1988bm}
~Gibbons G.~W.,
Sobolev's inequality, Jensen's theorem and the mass and entropy of the
universe, 1988
Nucl.\ Phys.\  B {\bf 310} 636.

\bibitem{Barrow:2003ni}
J.~D.~Barrow, G.~F.~R.~Ellis, R.~Maartens and C.~G.~Tsagas,
``On the stability of the Einstein static universe,''
Class.\ Quant.\ Grav.\  {\bf 20}, L155 (2003).
[gr-qc/0302094].



\bibitem{Mulryne:2005ef}
D.~J.~Mulryne, R.~Tavakol, J.~E.~Lidsey and G.~F.~R.~Ellis,
Phys. Rev. D \textbf{71} (2005), 123512
doi:10.1103/PhysRevD.71.123512
[arXiv:astro-ph/0502589 [astro-ph]].


\bibitem{Mukherjee:2005zt}
~Mukherjee S., Paul B.~C., Maharaj S.~D.  and ~Beesham A.,
Emergent universe in Starobinsky model,
arXiv:gr-qc/0505103.

\bibitem{Nunes:2005ra}
~Nunes N.~J.,
Inflation: A graceful entrance from loop quantum cosmology, 2005
Phys.\ Rev.\  D {\bf 72} 103510.



\bibitem{Mukherjee:2006ds}
~Mukherjee S., Paul B.~C., Dadhich N.~K., Maharaj S.~D. and Beesham
A.~,
Emergent universe with exotic matter, 2006
Class.\ Quant.\ Grav.\  {\bf 23} 6927.

\bibitem{Lidsey:2006md}
~Lidsey J.~E. and~Mulryne D.~J.,
A graceful entrance to braneworld inflation, 2006
Phys.\ Rev.\  D {\bf 73} 083508.


\bibitem{Banerjee:2007qi}
A.~Banerjee, T.~Bandyopadhyay and S.~Chakraborty,
Grav. Cosmol. \textbf{13} (2007), 290-292
[arXiv:0705.3933 [gr-qc]].

\bibitem{Banerjee:2007sg}
A.~Banerjee, T.~Bandyopadhyay and S.~Chakraborty,
Gen. Rel. Grav. \textbf{40} (2008), 1603-1607
doi:10.1007/s10714-007-0567-3
[arXiv:0711.4188 [gr-qc]].



\bibitem{Debnath:2008nu}
U.~Debnath,
Class. Quant. Grav. \textbf{25} (2008), 205019
doi:10.1088/0264-9381/25/20/205019
[arXiv:0808.2379 [gr-qc]].




\bibitem{Beesham:2009zw}
A.~Beesham, S.~V.~Chervon and S.~D.~Maharaj,
Class. Quant. Grav. \textbf{26} (2009), 075017
doi:10.1088/0264-9381/26/7/075017
[arXiv:0904.0773 [gr-qc]].

\bibitem{Kan:2009ws}
N.~Kan and K.~Shiraishi,
Prog. Theor. Phys. \textbf{121} (2009), 1035-1048
doi:10.1143/PTP.121.1035
[arXiv:0901.3879 [gr-qc]].

\bibitem{Paul:2009csp}
B.~C.~Paul and S.~Ghose,
Gen. Rel. Grav. \textbf{42} (2010), 795-812
doi:10.1007/s10714-009-0880-0
[arXiv:0809.4131 [hep-th]].

\bibitem{Wu:2009ah}
P.~Wu and H.~W.~Yu,
Phys. Rev. D \textbf{81} (2010), 103522
doi:10.1103/PhysRevD.81.103522
[arXiv:0909.2821 [gr-qc]].



\bibitem{Mukerji:2010zza}
S.~Mukerji and S.~Chakraborty,
Int. J. Theor. Phys. \textbf{49} (2010), 2446-2455
doi:10.1007/s10773-010-0430-2

\bibitem{Chakraborty:2010zzb}
S.~Chakraborty and U.~Debnath,
Int. J. Theor. Phys. \textbf{50} (2011), 80-87
doi:10.1007/s10773-010-0495-y

\bibitem{Mukerji:2010zz}
S.~Mukerji and S.~Chakraborty,
Astrophys. Space Sci. \textbf{331} (2011), 665-671
doi:10.1007/s10509-010-0456-1


\bibitem{Chattopadhyay:2011fp}
S.~Chattopadhyay and U.~Debnath,
Int. J. Mod. Phys. D \textbf{20} (2011), 1135-1152
doi:10.1142/S0218271811019293
[arXiv:1105.1091 [gr-qc]].

\bibitem{delCampo:2011mq}
S.~del Campo, E.~I.~Guendelman, A.~B.~Kaganovich, R.~Herrera and P.~Labrana,
Phys. Lett. B \textbf{699} (2011), 211-216
doi:10.1016/j.physletb.2011.03.061
[arXiv:1105.0651 [astro-ph.CO]].

\bibitem{Debnath:2011qi}
U.~Debnath and S.~Chakraborty,
Int. J. Theor. Phys. \textbf{50} (2011), 2892-2898
doi:10.1007/s10773-011-0789-8
[arXiv:1104.1673 [gr-qc]].

\bibitem{Paul:2011nw}
B.~C.~Paul, S.~Ghose and P.~Thakur,
Mon. Not. Roy. Astron. Soc. \textbf{413} (2011), 686
doi:10.1111/j.1365-2966.2010.18177.x
[arXiv:1101.1360 [astro-ph.CO]].



\bibitem{Rudra:2012mu}
P.~Rudra,
Mod. Phys. Lett. A \textbf{27} (2012), 1250189
doi:10.1142/S0217732312501891
[arXiv:1211.2047 [gr-qc]].

\bibitem{Cai:2012yf}
Y.~F.~Cai, M.~Li and X.~Zhang,
Phys. Lett. B \textbf{718} (2012), 248-254
doi:10.1016/j.physletb.2012.10.065
[arXiv:1209.3437 [hep-th]].


\bibitem{Singh:2012zzf}
C.~P.~Singh and V.~Singh,
Astrophys. Space Sci. \textbf{339} (2012), 101-109
doi:10.1007/s10509-012-0982-0

\bibitem{Liu:2012ww}
Z.~G.~Liu and Y.~S.~Piao,
Phys. Lett. B \textbf{718} (2013), 734-739
doi:10.1016/j.physletb.2012.11.068
[arXiv:1207.2568 [gr-qc]].

\bibitem{Labrana:2011np}
P.~Labrana,
Phys. Rev. D \textbf{86} (2012), 083524
doi:10.1103/PhysRevD.86.083524
[arXiv:1111.5360 [gr-qc]].



\bibitem{Cai:2013rna}
Y.~F.~Cai, Y.~Wan and X.~Zhang,
Phys. Lett. B \textbf{731} (2014), 217-226
doi:10.1016/j.physletb.2014.02.042
[arXiv:1312.0740 [hep-th]].



\bibitem{Chervon:2014tra}
S.~V.~Chervon, S.~D.~Maharaj, A.~Beesham and A.~S.~Kubasov,
Grav. Cosmol. \textbf{20} (2014), 176-181
doi:10.1134/S0202289314030074
[arXiv:1405.7219 [gr-qc]].

\bibitem{Chakraborty:2014ora}
S.~Chakraborty,
Phys. Lett. B \textbf{732} (2014), 81-84
doi:10.1016/j.physletb.2014.03.028
[arXiv:1403.5980 [gr-qc]].


\bibitem{Labrana:2013oca}
P.~Labra\~na,
Phys. Rev. D \textbf{91} (2015) no.8, 083534
doi:10.1103/PhysRevD.91.083534
[arXiv:1312.6877 [astro-ph.CO]].

\bibitem{Beesham:2014tja}
A.~Beesham, S.~V.~Chervon, S.~D.~Maharaj and A.~S.~Kubasov,
Int. J. Theor. Phys. \textbf{54} (2015), 884-895
doi:10.1007/s10773-014-2284-5

\bibitem{Labrana:2014yta}
P.~Labra\~na,
AIP Conf. Proc. \textbf{1606} (2015) no.1, 38-47
doi:10.1063/1.4891114
[arXiv:1406.0922 [astro-ph.CO]].


\bibitem{Paul:2015eja}
B.~C.~Paul and A.~Majumdar,
Class. Quant. Grav. \textbf{32} (2015) no.11, 115001
doi:10.1088/0264-9381/32/11/115001
[arXiv:1503.08284 [gr-qc]].



\bibitem{Guendelman:2015uca}
E.~Guendelman, R.~Herrera, P.~Labrana, E.~Nissimov and S.~Pacheva,
Astron. Nachr. \textbf{336} (2015) no.8/9, 810-814
doi:10.1002/asna.201512221
[arXiv:1507.08878 [hep-th]].



\bibitem{delCampo:2015yfa}
S.~del Campo, E.~I.~Guendelman, R.~Herrera and P.~Labra\~na,
JCAP \textbf{08} (2016), 049
doi:10.1088/1475-7516/2016/08/049
[arXiv:1508.03330 [gr-qc]].


\bibitem{Heydarzade:2015dba}
Y.~Heydarzade, H.~Hadi, F.~Darabi and A.~Sheykhi,
Eur. Phys. J. C \textbf{76} (2016) no.6, 323
doi:10.1140/epjc/s10052-016-4162-1
[arXiv:1506.02388 [gr-qc]].


\bibitem{Dutta:2015fha}
J.~Dutta, S.~Haldar and S.~Chakraborty,
Astrophys. Space Sci. \textbf{361} (2016) no.1, 21
doi:10.1007/s10509-015-2607-x
[arXiv:1505.01263 [gr-qc]].


\bibitem{Singh:2015nqn}
J.~K.~Singh and S.~Rani,
Int. J. Theor. Phys. \textbf{55} (2016) no.1, 232-240
doi:10.1007/s10773-015-2655-6


\bibitem{Khodadi:2015fav}
M.~Khodadi, Y.~Heydarzade, F.~Darabi and E.~N.~Saridakis,
Phys. Rev. D \textbf{93} (2016) no.12, 124019
doi:10.1103/PhysRevD.93.124019
[arXiv:1512.08674 [gr-qc]].


\bibitem{Guendelman:2014bva}
E.~Guendelman, R.~Herrera, P.~Labrana, E.~Nissimov and S.~Pacheva,
Gen. Rel. Grav. \textbf{47} (2015) no.2, 10
doi:10.1007/s10714-015-1852-1
[arXiv:1408.5344 [gr-qc]].



\bibitem{Gangopadhyay:2014jfa}
S.~Gangopadhyay, A.~Saha and S.~Mukherjee,
Int. J. Theor. Phys. \textbf{55} (2016) no.10, 4445-4452
doi:10.1007/s10773-016-3067-y
[arXiv:1412.2567 [gr-qc]].




\bibitem{Hadi:2016uaw}
H.~Hadi, Y.~Heydarzade and F.~Darabi,
[arXiv:1609.05629 [gr-qc]].

\bibitem{Sharif:2016hcx}
M.~Sharif and A.~Sarwar,
Mod. Phys. Lett. A \textbf{31} (2016) no.22, 1650129
doi:10.1142/S0217732316501297

\bibitem{Bhattacharya:2016env}
S.~Bhattacharya and S.~Chakraborty,
Class. Quant. Grav. \textbf{33} (2016) no.3, 035013
doi:10.1088/0264-9381/33/3/035013
[arXiv:1601.03816 [gr-qc]].

\bibitem{Labrana:2016jmm}
P.~Labra\~na,
J. Phys. Conf. Ser. \textbf{720} (2016), 012016
doi:10.1088/1742-6596/720/1/012016

\bibitem{Rios:2016trs}
C.~R\'\i{}os, P.~Labra\~na and A.~Cid,
J. Phys. Conf. Ser. \textbf{720} (2016), 012008
doi:10.1088/1742-6596/720/1/012008

\bibitem{Alesci:2016xqa}
E.~Alesci, G.~Botta, F.~Cianfrani and S.~Liberati,
Phys. Rev. D \textbf{96} (2017) no.4, 046008
doi:10.1103/PhysRevD.96.046008
[arXiv:1612.07116 [gr-qc]].



\bibitem{Debnath:2017xcu}
P.~S.~Debnath and B.~C.~Paul,
Mod. Phys. Lett. A \textbf{32} (2017) no.39, 1750216
doi:10.1142/S0217732317502169



\bibitem{Atazadeh:2017xwe}
M.~Atazadeh,
doi:10.1142/9789813226609\_0104


\bibitem{Marachlian:2015cia}
E.~Marachlian, I.~E.~S\'anchez G. and O.~P.~Santill\'an,
Mod. Phys. Lett. A \textbf{32} (2017) no.28, 1750152
doi:10.1142/S0217732317501528
[arXiv:1508.05083 [gr-qc]].

\bibitem{Ghosh:2016fue}
S.~Ghosh and S.~Gangopadhyay,
Mod. Phys. Lett. A \textbf{32} (2017) no.16, 1750089
doi:10.1142/S0217732317500894
[arXiv:1609.09779 [gr-qc]].



\bibitem{Martineau:2018isp}
K.~Martineau and A.~Barrau,
Universe \textbf{4} (2018) no.12, 149
doi:10.3390/universe4120149
[arXiv:1812.05522 [gr-qc]].


\bibitem{Labrana:2018ogv}
P.~Labra\~na, S.~del Campo, R.~Herrera and E.~Guendelman,
J. Phys. Conf. Ser. \textbf{1043} (2018) no.1, 012026
doi:10.1088/1742-6596/1043/1/012026


\bibitem{Ghose:2017xop}
S.~Ghose,
Pramana \textbf{90} (2018) no.3, 43
doi:10.1007/s12043-018-1535-z
[arXiv:1712.01556 [gr-qc]].

\bibitem{Khodadi:2016gyw}
M.~Khodadi, K.~Nozari and E.~N.~Saridakis,
Class. Quant. Grav. \textbf{35} (2018) no.1, 015010
doi:10.1088/1361-6382/aa95aa
[arXiv:1612.09254 [gr-qc]].


\bibitem{Debnath:2019mkz}
P.~S.~Debnath,
Int. J. Geom. Meth. Mod. Phys. \textbf{16} (2019) no.11, 1950169
doi:10.1142/S021988781950169X

\bibitem{Paul:2019oxo}
B.~C.~Paul and A.~Chanda,
Gen. Rel. Grav. \textbf{51} (2019) no.6, 71
doi:10.1007/s10714-019-2551-0

\bibitem{Li:2019laq}
S.~L.~Li, H.~L\"u, H.~Wei, P.~Wu and H.~Yu,
Phys. Rev. D \textbf{99} (2019) no.10, 104057
doi:10.1103/PhysRevD.99.104057
[arXiv:1903.03940 [gr-qc]].

\bibitem{Olmedo:2018ohq}
J.~Olmedo and E.~Alesci,
JCAP \textbf{04} (2019), 030
doi:10.1088/1475-7516/2019/04/030
[arXiv:1811.04327 [gr-qc]].

\bibitem{Cai:2018ebs}
R.~G.~Cai, S.~Khimphun, B.~H.~Lee, S.~Sun, G.~Tumurtushaa and Y.~L.~Zhang,
Phys. Dark Univ. \textbf{26} (2019), 100387
doi:10.1016/j.dark.2019.100387
[arXiv:1812.11105 [hep-th]].

  


\bibitem{Paul:2020bje}
B.~C.~Paul, S.~D.~Maharaj and A.~Beesham,
[arXiv:2008.00169 [astro-ph.CO]].


\bibitem{Debnath:2020bno}
P.~S.~Debnath and B.~C.~Paul,
Int. J. Geom. Meth. Mod. Phys. \textbf{17} (2020) no.07, 2050102
doi:10.1142/S0219887820501029


  \bibitem{Paul:2021lvb}
B.~C.~Paul,
Eur. Phys. J. C \textbf{81} (2021) no.8, 776
doi:10.1140/epjc/s10052-021-09562-2


\bibitem{Bengochea:2021jvt}
G.~R.~Bengochea, M.~P.~Piccirilli and G.~Le\'on,
[arXiv:2108.01472 [gr-qc]].


\bibitem{Thakur:2021ufp}
P.~Thakur,
doi:10.1007/s12648-021-02018-z


\bibitem{Thakur:2021ryp}
P.~Thakur,
Pramana \textbf{95} (2021) no.2, 91
doi:10.1007/s12043-021-02124-x


\bibitem{Hamil:2021ivf}
B.~Hamil, M.~Merad and T.~Birkandan,
Rev. Mex. Fis. \textbf{67} (2021) no.2, 219-225
doi:10.31349/RevMexFis.67.219



\bibitem{Ilyas:2020zcb}
A.~Ilyas, M.~Zhu, Y.~Zheng and Y.~F.~Cai,
JHEP \textbf{01} (2021), 141
doi:10.1007/JHEP01(2021)141
[arXiv:2009.10351 [gr-qc]].




\bibitem{delCampo:2007mp}
S.~del Campo, R.~Herrera and P.~Labrana,
``Emergent universe in a Jordan-Brans-Dicke theory,''
JCAP {\bf 0711}  030 (2007).
[arXiv:0711.1559 [gr-qc]].


\bibitem{delCampo:2009kp}
S.~del Campo, R.~Herrera and P.~Labrana,
``On the Stability of Jordan-Brans-Dicke Static Universe,''
JCAP {\bf 0907} (2009) 006.
[arXiv:0905.0614 [gr-qc]].

\bibitem{Huang:2014fia}
  H.~Huang, P.~Wu and H.~Yu,
  Phys.\ Rev.\ D {\bf 89}, no. 10, 103521 (2014).

\bibitem{Labrana:2018bkw}
P.~Labrana and H.~Cossio,
Eur. Phys. J. C \textbf{79} (2019) no.4, 303
doi:10.1140/epjc/s10052-019-6811-7
[arXiv:1808.09291 [gr-qc]].

\bibitem{Cossio:2018xor}
H.~Cossio and P.~Labra\~na,
J. Phys. Conf. Ser. \textbf{1043} (2018) no.1, 012021
doi:10.1088/1742-6596/1043/1/012021




\bibitem{Jbd}  Jordan P.,  The present state of Dirac's cosmological
hypothesis, 1959 Z.Phys. {\bf 157} 112;
  Brans C.~ and ~Dicke R.~H.,
  Mach's principle and a relativistic theory of gravitation, 1961
  Phys.\ Rev.\  {\bf 124} 925.





\bibitem{Freund:1982pg}
  ~Freund P.~G.~O.,
  Kaluza-Klein cosmologies, 1982
  Nucl.\ Phys.\  B {\bf 209} 146.

\bibitem{Appelquist:1987nr}
  Appelquist T., Chodos A. and ~Freund P.~G.~O.,
%
\textit{Modern Kaluza-Klein theories} ( 1987 Addison-Wesley,
Redwood City).
%

\bibitem{Fradkin:1984pq}
  ~Fradkin E.~S. and Tseytlin A.~A.,
  Effective field theory from quantized strings, 1985
  Phys.\ Lett.\  B {\bf 158} 316.

\bibitem{Fradkin:1985ys}
 ~Fradkin E.~S. and ~Tseytlin A.~A.,
  Quantum string theory effective action, 1985
  Nucl.\ Phys.\  B {\bf 261} 1.

\bibitem{Callan:1985ia}
  Callan C.~G., Martinec E.~J., Perry M.~J.~ and ~Friedan D.,
  Strings in background fields, 1985
  Nucl.\ Phys.\  B {\bf 262} 593.

\bibitem{Callan:1986jb}
  CallanC.~G., Klebanov I.~R.~ and Perry M.~J.~,
  String theory effective actions, 1986
  Nucl.\ Phys.\  B {\bf 278} 78.

\bibitem{Green:1987sp}
  ~Green M.~B., ~Schwarz J.~H. and ~Witten E., \textit{Superstring theory}
 (1987 Cambridge, Uk: Univ. Pr., Cambridge Monographs On
Mathematical Physics).


\bibitem{Lesnefsky:2022fen}
J.~E.~Lesnefsky, D.~A.~Easson and P.~C.~W.~Davies,
``Past-completeness of inflationary spacetimes,''
Phys. Rev. D \textbf{107} (2023) no.4, 044024
doi:10.1103/PhysRevD.107.044024
[arXiv:2207.00955 [gr-qc]].

\bibitem{Kinney:2023urn}
W.~H.~Kinney, S.~Maity and L.~Sriramkumar,
``Borde-Guth-Vilenkin theorem in extended de Sitter spaces,''
Phys. Rev. D \textbf{109} (2024) no.4, 043519
doi:10.1103/PhysRevD.109.043519
[arXiv:2307.10958 [gr-qc]].

\bibitem{Geshnizjani:2023hyd}
G.~Geshnizjani, E.~Ling and J.~Quintin,
``On the initial singularity and extendibility of flat quasi-de Sitter spacetimes,''
JHEP \textbf{10} (2023), 182
doi:10.1007/JHEP10(2023)182
[arXiv:2305.01676 [gr-qc]].

\bibitem{Easson:2024fzn}
D.~A.~Easson and J.~E.~Lesnefsky,
``Inflationary resolution of the initial singularity,''
[arXiv:2402.13031 [hep-th]].

\bibitem{Easson:2024vpp}
D.~A.~Easson and J.~E.~Lesnefsky,
``Eternal universes,''
Phys. Rev. D \textbf{112} (2025), 063545
doi:10.1103/PhysRevD.112.063545
[arXiv:2404.03016 [hep-th]].

\bibitem{Moghtaderi:2025cns}
E.~Moghtaderi, B.~R.~Hull, J.~Quintin and G.~Geshnizjani,
``How much NEC breaking can the Universe endure?,''
Phys. Rev. D \textbf{111} (2025) no.12, 123552
doi:10.1103/PhysRevD.111.123552
[arXiv:2503.19955 [gr-qc]].

\bibitem{Ferreira:2025zsy}
A.~L.~Ferreira~Jr, N.~Pinto-Neto and V.~N.~Xavier,
``On the bouncing completion of eternal inflation,''
[arXiv:2504.01224 [gr-qc]].

\bibitem{Burwig:2025lyq}
N.~L.~Burwig and D.~A.~Easson,
``Open case for a closed universe,''
Phys. Rev. D (2026), doi:10.1103/mn3v-myzc
[arXiv:2510.13971 [hep-th]].

\bibitem{Kinney:2026abc}
W.~H.~Kinney,
``Geodesic Completeness in General Cosmological Scenarios,''
[arXiv:2604.20809 [gr-qc]].

\bibitem{Easson:2026abc}
D.~A.~Easson,
``Geodesically Complete Curvature-Bounce Inflation,''
[arXiv:2604.27103 [gr-qc]].

\bibitem{Burwig:2026abc}
N.~L.~Burwig and D.~A.~Easson,
``Affine ANEC selects the closed FRW branch for geodesically complete cosmology,''
[arXiv:2605.18965 [gr-qc]].


\bibitem{Green:2012oqa}
M.~B.~Green, J.~H.~Schwarz and E.~Witten,
``Superstring Theory Vol. 1 : 25th Anniversary Edition,''
doi:10.1017/CBO9781139248563


\bibitem{Landau:1982dva}
L.~D.~Landau and E.~M.~Lifshitz,
\emph{The Classical Theory of Fields}, 4th revised English edition,
Course of Theoretical Physics Vol.~2 (Butterworth-Heinemann, Oxford, 1980).


\bibitem{Vilenkin:1987kf}
  A.~Vilenkin,
  Phys.\ Rev.\ D {\bf 37}, 888 (1988).
  doi:10.1103/PhysRevD.37.888.

\bibitem{Bertotti:2003rm}
B.~Bertotti, L.~Iess and P.~Tortora,
``A test of general relativity using radio links with the Cassini spacecraft,''
Nature \textbf{425} (2003), 374-376
doi:10.1038/nature01997

\bibitem{Will:2014kxa}
C.~M.~Will,
``The Confrontation between General Relativity and Experiment,''
Living Rev. Rel. \textbf{17} (2014), 4
doi:10.12942/lrr-2014-4
[arXiv:1403.7377 [gr-qc]].

\bibitem{Perivolaropoulos:2009ak}
L.~Perivolaropoulos,
``PPN Parameter gamma and Solar System Constraints of Massive Brans-Dicke Theories,''
Phys. Rev. D \textbf{81} (2010), 047501
doi:10.1103/PhysRevD.81.047501
[arXiv:0911.3401 [gr-qc]].

\bibitem{Will} B.S. DeWitt, Quantum Theory of Gravity. 1. The Canonical Theory, Phys. Rev. \textbf{160}, 1113 (1967).

\bibitem{Ratra:1987rm}
B.~Ratra and P.~J.~E.~Peebles,
``Cosmological Consequences of a Rolling Homogeneous Scalar Field,''
Phys. Rev. D \textbf{37} (1988), 3406
doi:10.1103/PhysRevD.37.3406

\bibitem{Binetruy:1998rz}
P.~Binetruy,
``Models of dynamical supersymmetry breaking and quintessence,''
Phys. Rev. D \textbf{60} (1999), 063502
doi:10.1103/PhysRevD.60.063502
[arXiv:hep-ph/9810553 [hep-ph]].


\end{thebibliography}
\end{document}